\documentstyle [12pt,eqsecnum,aps,amsfonts] {revtex}
\input epsf
\newcommand{\beq}{\begin{equation}}
\newcommand{\eeq}{\end{equation}}
\newcommand{\beqs}{\begin{eqnarray}}
\newcommand{\eeqs}{\end{eqnarray}}

\begin{document}
\draft

\baselineskip 6.0mm

\title{Ground State Entropy of Potts Antiferromagnets on Homeomorphic
Families of Strip Graphs}

\author{Robert Shrock\thanks{email: shrock@insti.physics.sunysb.edu}
\and Shan-Ho Tsai\thanks{email: tsai@insti.physics.sunysb.edu}}

\address{
Institute for Theoretical Physics  \\
State University of New York       \\
Stony Brook, N. Y. 11794-3840}

\maketitle

\vspace{10mm}

\begin{abstract}

We present exact calculations of the zero-temperature partition function,
and ground-state degeneracy (per site), $W$, for the $q$-state Potts 
antiferromagnet on a variety of homeomorphic families of planar strip
graphs $G = (Ch)_{k_1,k_2,\Sigma,k,m}$, where $k_1$, $k_2$, $\Sigma$, and $k$ 
describe the homeomorphic structure, and $m$ denotes the length of the strip. 
Several different ways of taking the total number of vertices to infinity, by
sending (i) $m \to \infty$ with $k_1$, $k_2$, and $k$ fixed; (ii) 
$k_1$ and/or $k_2 \to \infty$ with $m$, and $k$ fixed; and (iii) 
$k \to \infty$ with $m$ and $p=k_1+k_2$ fixed are studied and the respective 
loci of points ${\cal B}$ where $W$ is nonanalytic in the complex $q$
plane are determined.  The ${\cal B}$'s for limit (i) are comprised of arcs
which do not enclose regions in the $q$ plane and, for many values of $p$ and 
$k$, include support for $Re(q) < 0$.  The ${\cal B}$ for limits (ii) and (iii)
is the unit circle $|q-1|=1$. 

\end{abstract}


\vspace{16mm}

\pagestyle{empty}
\newpage

\pagestyle{plain}
\pagenumbering{arabic}
\renewcommand{\thefootnote}{\arabic{footnote}}
\setcounter{footnote}{0}

\section{Introduction}

   This paper continues our study of nonzero ground state entropy,
$S_0(\{G\},q) \ne 0$, i.e., ground state degeneracy (per site)
$W(\{G\},q) > 1$, where $S_0 = k_B \ln W$, in $q$-state Potts 
antiferromagnets \cite{potts,wurev} on various lattices and, more generally, 
families of graphs $\{G\}$. There is an interesting connection with graph 
theory here, since the zero-temperature partition function of the 
above-mentioned $q$-state Potts antiferromagnet (PAF) on a graph
$G$ satisfies $Z(G,q,T=0)_{PAF}=P(G,q)$, where $P(G,q)$ is the chromatic
polynomial expressing the number of ways of coloring the vertices of the graph
$G$ \cite{graphdef} with $q$ colors such that no two adjacent vertices have 
the same color \cite{birk}--\cite{biggsbook}.  Thus,
\beq
W([\lim_{n \to \infty} G \ ],q) = \lim_{n \to \infty} P(G,q)^{1/n}
\label{w}
\eeq
where $n=v(G)$ is the number of vertices of $G$.
An example of a substance exhibiting nonzero ground
state entropy is ice \cite{lp,liebwu}.  Just as complex analysis
provides deeper insights into real analysis in mathematics, the generalization
from $q \in {\mathbb Z}_+$, to $q \in {\mathbb C}$ yields a deeper
understanding of the behavior of $W(\{G\},q)$ for physical (positive integral)
$q$.  In general, $W(\{G\},q)$ is an analytic function in the $q$ plane
except along a certain continuous locus of points, which we denote ${\cal B}$.
In the limit as $n \to \infty$, the locus ${\cal B}$ forms by means of a
coalescence of a subset of the zeros of $P(G,q)$ (called chromatic zeros of
$G$) \cite{early}. 

In a series of papers \cite{p3afhc}-\cite{wa23}
we have calculated and analyzed $W(\{G\},q)$, both for 
physical values of $q$ (via rigorous upper and lower bounds, large--$q$ series
calculations, and Monte Carlo measurements) and for the generalization to
complex values of $q$.  In the present work we construct a variety of
homeomorphic families of graphs and give exact calculations of the chromatic
polynomials, $W$ functions, and resultant nonanalytic loci ${\cal B}$.  We also
compare these loci with zeros of the chromatic polynomials for graphs with
reasonably large number of vertices.  Our results extend our previous 
study, with M. Ro\v{c}ek, of open (planar) 
strip graphs of lattices \cite{strip} via 
homeomorphic expansion, as will be discussed in detail below.  The families 
of graphs that we will construct and study depend on a set of (positive
integer) parameters, including homeomorphic expansion indices $k_1$, $k_2$, and
$k$; the number $m$ of longitudinal edges along the strip; and a vector 
$\Sigma=(\sigma_1,...,\sigma_m)$ that describes the allocation of the indices 
$k_1$ and $k_2$ in the homeomorphic expansions of the various longitudinal
edges. As we shall show, the chromatic polynomial for such a graph is 
independent of $\Sigma$ and only depends on the first two homeomorphic 
expansion indices $k_1$ and $k_2$ through their sum, $p=k_1+k_2$.  
Since the number of vertices $n=v$ is a linear function of $p$, $k$,
and $m$, there are several ways of producing the limit $n \to \infty$ 
($L$ denotes limit):
\beq
L_m : \ m \to \infty \quad {\rm with} \quad k \ , p \quad {\rm fixed}
\label{minf}
\eeq
\beq
L_p: \ p \to \infty \quad {\rm with} \quad m \ , k \quad {\rm fixed}
\label{pinf}
\eeq
and 
\beq
L_k: \ k \to \infty \quad {\rm with} \quad p, \ m \quad {\rm fixed}
\label{kinf}
\eeq
We shall concentrate on the $L_m$ limit here, i.e., the limit of an infinitely
long, finite-width planar
strip, since this is the most interesting of the three
limits and since it constitutes a homeomorphic generalization of the previous 
study in Ref. \cite{strip}.  It is natural to begin with the simplest such 
strips, of minimal nontrivial width for this homeomorphic expansion, and we
shall do so; as will be seen, the results already exhibit considerable richness
and complexity.  We shall also briefly discuss the other two 
limits, $L_p$ and $L_k$. 

For our results below, it will be convenient to define the polynomial 
\beq
D_n(q) = \frac{P(C_n,q)}{q(q-1)} = a^{n-2}\sum_{j=0}^{n-2}(-a)^{-j} =
\sum_{s=0}^{n-2}(-1)^s {{n-1}\choose {s}} q^{n-2-s}
\label{dk}
\eeq
where
\beq
a=q-1
\label{a}
\eeq
and $P(C_n,q)$ is the chromatic polynomial for the circuit graph with $n$
vertices:
\beq
P(C_n,q) = a^n + (-1)^na 
\label{pck}
\eeq
The terms ``edge'' and ``bond'' will be used synonymously. 

The organization of the paper is as follows.  In section 2 we discuss 
the connections between the structure of the chromatic polynomials, their
generating functions, and recursion relations that these polynomials satisfy.
In section 3 we construct and classify the homeomorphic expansions of strip
(chain) graphs.  Section 4 contains our calculations of the generating function
and hence chromatic polynomials for these homeomorphic families of strip 
graphs.  This section also includes the calculation of the $W$ function for the
$L_m$ limit.  In section 5 we present explicit results for the $L_m$ limit for
various $p$ and $k$ values, and in section 6 we discuss and explain general
properties of the analytic structure of $W$ for this limit.  Section 7 presents
analogous results for the $L_p$ and $L_k$ limits, together with a comparison of
the three limits.  Some concluding remarks are given in section 8.  
Useful properties of $D_k$ are listed in Appendix 1. 

\section{Connections Between Structure of Chromatic Polynomial, Generating
Function, and Recursion Relation}

\subsection{Generating Function and Chromatic Polynomial} 

   A general form for the chromatic polynomial of an $n$-vertex graph $G$ is
\beq
P(G,q) =  c_0(q) + \sum_{j=1}^{N_a} c_j(q)(a_j(q))^{t_j n}
\label{pgsum}
\eeq
where $c_j(q)$ and $a_j(q)$ are certain functions of $q$. Here the $a_j(q)$ and
$c_{j \ne 0}(q)$ are independent of $n$, while $c_0(q)$ may contain
$n$-dependent terms, such as $(-1)^n$, but does not grow with $n$ like
$(const.)^n$ with $|const.| > 1$.  As before \cite{w}, we define a term 
$a_\ell(q)$ as ``leading'' if it dominates
the $n \to \infty$ limit of $P(G,q)$; in particular, if $N_a \ge 2$, then it
satisfies $|a_\ell(q)| \ge 1$ and $|a_\ell(q)| > |a_j(q)|$ for $j \ne \ell$, so
that $|W|=|a_\ell|^{t_j}$.  The locus ${\cal B}$ occurs where
there is a nonanalytic change in $W$ as the leading terms $a_\ell$
in eq. (\ref{pgsum}) changes.  Note that the $c_0(q)$ term may be absent. 
For some families of graphs, the $c_j(q)$ and $a_j(q)$ are polynomials.
However, there are also many families of graphs for which $c_j(q)$ and 
$a_j(q)$ are not polynomial, but instead, are algebraic, functions of $q$; for
these families the property that the chromatic polynomial is, in fact, a
polynomial of $q$, is not manifest from the expression (\ref{pgsum}).  A number
of families of this latter type were analyzed in Refs. \cite{strip,wa23}; for
these we used a generating function method to calculate the chromatic
polynomials.  

Here we discuss the connection between our generating function method and the 
form (\ref{pgsum}).  Let a graph $G_1$ be constructed by a certain number of 
successive additions of a repeating graphical subunit $H$ to an initial 
subgraph $I$, where $I$ may be the same as $H$. This procedure is similar to
the method described in Ref. \cite{strip} \cite{label}.  However, graphs 
$G_1$ constructed in this fashion are general recursive families of graphs 
and need not be strip graphs.  The graphs typically also depend also on other
parameters, which will be denoted $\{k\}$ but will be suppressed in the
notation in this section; for the applications in the present
paper these will be certain homeomorphic expansion indices, as will be
described in greater detail below.  The total number of vertices is given by 
\beq
n = v(G_m) = \kappa_1 m + \kappa_0
\label{nm}
\eeq
where $\kappa_1$ and $\kappa_0$ depend on $\{k\}$.  
Applying the deletion-contraction theorem \cite{thm} to the graph $G_1$, one 
obtains a finite set of linear equations, with $m$-independent coefficients, 
relating $P((G_1)_m,q)$, $P((G_1)_{m-1},q)$, $P((G_f)_m,q)$, and possibly 
$P((G_f)_{m-1},q)$, where $G_f$, with $f=2,..,n_f$, denotes other recursively
defined families of graphs. 
As was the case for the strip graphs discussed in Ref. \cite{strip}, 
the set of equations is linear because the
deletion-contraction theorem is a linear equation among chromatic polynomials. 
The number $n_f$ of families of graphs involved in the set of linear equations
is finite because the repeating subgraph $H$ is a finite graph. This set of 
linear equations can be solved by using generating functions of the form 
\beq
\Gamma(G_f,q,x) = \sum_{m=0}^{\infty}P((G_f)_m,q)x^m, \quad {\rm where} \quad 
f=1,2,..,n_f
\label{gamma}
\eeq
where $x$ is an auxiliary variable. 
We find that the generating functions $\Gamma(G_f,q,x)$ are rational functions 
of the form 
\beq
\Gamma(G_f,q,x) = \frac{{\cal N}(G_f,q,x)}{{\cal D}(G_f,q,x)}
\label{gammagen}
\eeq
with
\beq
{\cal N}(G_f,q,x) = \sum_{j=0}^{d_{\cal N}} A_{f,j}(q) x^j
\label{n}
\eeq
and
\beq
{\cal D}(G_f,q,x) = 1 + \sum_{j=1}^{d_{\cal D}} b_{f,j}(q) x^j 
\label{d}
\eeq
where the $A_{f,i}$ and $b_{f,i}$ are polynomials in $q$ that depend on the 
family of graphs $G_f$, and the degrees of the numerator and denominator,
as polynomials in the auxiliary variable $x$, are denoted
\beq
d_{\cal N}(G_f) = deg_x({\cal N}(G_f))
\label{degn}
\eeq
\beq
d_{\cal D}(G_f) = deg_x({\cal D}(G_f))
\label{degd}
\eeq
Let us write the denominator 
of the generating function $\Gamma(G_f,q,x)$ in factorized form
\beq
{\cal D}(G_f,q,x) = \prod_{j=1}^{d_{\cal D}}(1-\lambda_{f,j}(q)x)
\label{lambdaform}
\eeq
Henceforth, for brevity, we suppress the $G_f$-dependence in the notation. 
 Using the partial fraction identity
\beq
\prod_{j=1}^{d_{\cal D}} \frac{1}{(1-\lambda_jx)} = \sum_{j=1}^{d_{\cal D}}
\Biggl [ \frac{\lambda_j^{d_{\cal D}-1}}{(1-\lambda_jx)}
\Bigl [ \prod_{1 \le i \le d_{\cal D}; \ i \ne j}
\frac{1}{(\lambda_j-\lambda_i)} 
\Bigr ] \Biggr ]
\label{denproduct}
\eeq
with the identity, for each $j$, 
\beq
\frac{1}{1-\lambda_j x}=\sum_{m=0}^{\infty}\lambda_j^m x^m
\label{expansionx}
\eeq
in eq. (\ref{gammagen}), with (\ref{n}), we obtain the result
\beq
P(G_m,q) = \sum_{s=0}^{d_{\cal N}}A_s F_{m-s}
\label{chrompgsum}
\eeq
where
\beq
F_r = \sum_{j=1}^{d_{\cal D}}\Biggl [ (\lambda_j)^{r+d_{\cal D}-1}
\Bigl [ \prod_{1 \le i \le d_{\cal D}; \ i \ne j} 
\frac{1}{(\lambda_j-\lambda_i)} \Bigr ] \Biggr ]
\label{afr}
\eeq
with the restriction that  $d_{\cal N} \ < \ d_{\cal D}$, as is true for all
families of graphs considered.  Equivalently, 
\beq
P(G_m,q) = \sum_{j=1}^{d_{\cal D}} \Biggl [ \sum_{s=0}^{d_{\cal N}}
A_s \lambda_j^{d_{\cal D}-s-1} \Biggr ] 
\Biggl [ \prod_{1 \le i \le d_{\cal D}; i \ne j}
\frac{1}{(\lambda_j-\lambda_i)} \Biggr ] \lambda_j^m
\label{chrompgsumlam}
\eeq
For example, for $d_{\cal N}=1$ and $d_{\cal D}=2$, as was the case for a 
number of strip graphs studied in Ref. \cite{strip},  
eq. (\ref{chrompgsumlam}) reduces to
\beq
P(G_m,q) = \frac{(A_0 \lambda_1 + A_1)}
{(\lambda_1-\lambda_2)}\lambda_1^m + 
 \frac{(A_0 \lambda_2 + A_1)}
{(\lambda_2-\lambda_1)}\lambda_2^m
\label{pgsumkmax2}
\eeq
For the more complicated case $d_{\cal N}=2$ and $d_{\cal D}=3$, 
eq. (\ref{chrompgsumlam}) reduces to
\beqs
P(G_m,q) &=& \frac{(A_0\lambda_1^2+A_1\lambda_1+A_2)}
{(\lambda_1-\lambda_2)(\lambda_1-\lambda_3)}\lambda_1^m +
\frac{(A_0\lambda_2^2+A_1\lambda_2+A_2)}
{(\lambda_2-\lambda_1)(\lambda_2-\lambda_3)}\lambda_2^m +
\cr\cr
& & \frac{(A_0\lambda_3^2+A_1\lambda_3+A_2)}
{(\lambda_3-\lambda_1)(\lambda_3-\lambda_2)}\lambda_3^m
\label{pgsumkmax3}
\eeqs
and so forth for higher values of $d_{\cal N}$ and $d_{\cal D}$. 

Finally, we use eq. (\ref{nm}) to express eq. (\ref{chrompgsum}) in the form
(\ref{pgsum}): 
\beq
P(G_m,q) = \sum_{j=1}^{N_a} c_j (a_j)^{t_j n}
\label{pgsumgm}
\eeq
where 
\beq
N_a = d_{\cal D}
\label{naeqdegden}
\eeq
\beq
c_j =  \Biggl [ \sum_{s=0}^{d_{\cal N}}
A_s \lambda_j^{d_{\cal D}-s-1} \Biggr ]
\Biggl [ \prod_{1 \le i \le d_{\cal D}; i \ne j}
\frac{1}{(\lambda_j-\lambda_i)} \Biggr ]\lambda_j^{-\kappa_0/\kappa_1}
\label{cjgm}
\eeq
\beq
a_j = \lambda_j \quad , \quad j = 1,...,d_{\cal D} 
\label{ajlamjgm}
\eeq
and
\beq
t_j = \frac{1}{\kappa_1} \quad \forall \quad j
\label{tjkappa1}
\eeq
The advantage of the formulas for the chromatic polynomial given in eq. 
(\ref{chrompgsumlam}) or equivalently eqs. (\ref{pgsumgm})--(\ref{tjkappa1}) 
is that they show manifestly the role of the $\lambda_j$'s as the terms $a_j$,
and thus show that the equation for the continuous locus of points ${\cal B}$ 
where $W$ is nonanalytic is given by the degeneracy of leading terms 
\beq
|\lambda_{max}| =|\lambda_{max}'|
\label{lambdamax}
\eeq
where ``leading'' or maximal $\lambda_j$ is defined in the same way as the 
definition given above of the leading term $a_j$.  However, the 
fact that the chromatic polynomial is, indeed, a polynomial in $q$, is not
always manifest in the formulas (\ref{chrompgsumlam}), 
(\ref{pgsumgm})--(\ref{tjkappa1}) since the $c_j$'s and $\lambda_j$'s may be
algebraic, but nonpolynomial, functions of $q$. 

\subsection{Generating Function and Recursion Relation}

Writing eq. (\ref{gammagen}) as ${\cal D}\Gamma = {\cal N}$ and substituting
eq. (\ref{gamma}), we obtain
\beq
(1+\sum_{s=1}^{d_{\cal D}} b_s x^s)\sum_{m=0}^\infty P(G_m,q)x^m = 
\sum_{j=0}^{d_{\cal N}} A_j x^j
\label{dgn}
\eeq
which can be rewritten in the form 
\beq
\sum_{m=d_{\cal D}}^{\infty}x^m \Biggl [ \sum_{s=0}^{d_{\cal D}} b_s
P(G_{m-s},q) \Biggr ] + \sum_{m=0}^{d_{\cal D}-1}x^m \Biggl [ \sum_{s=0}^m 
b_s P(G_{m-s},q) \Biggr ] = \sum_{j=0}^{d_{\cal N}} A_j x^j
\label{recureq}
\eeq
Equating the coefficients of the same powers of $x$ on either side of this
equation yields the recursion relations 
$\sum_{s=0}^j b_s P(G_{j-s},q) = A_j$ for $j < d_{\cal D}$ and 
$\sum_{s=0}^{d_{\cal D}} b_s P(G_{j-s},q) = A_j$ for $j \ge d_{\cal D}$.  These
can be expressed as the single recursion relation 
\beq
\sum_{s=0}^{min(j,d_{\cal D})} b_s P(G_{j-s},q) = A_j
\label{recursionrel}
\eeq
where $min(x,y)$ denotes the minimum of $x$ and $y$ \cite{readrec}. 
This recursion relation holds for arbitrary $d_{\cal N}$ and $d_{\cal D}$ and 
involves only one family of graphs $G$. 

\section{Construction of Homeomorphic Expansions of Families of Strip Graphs}

We now study homeomorphic expansions of families of lattice strip graphs,
including the limit of infinite length (with finite width).  Homeomorphic 
classes of graphs have been of continuing interest in graph theory 
\cite{biggsbook}; for some recent theorems in a different direction from that
of the present work, see Ref. \cite{rw}.  We have previously used the method
of homeomorphic expansion to generate a large variety of families of graphs 
with noncompact $W$ boundaries ${\cal B}$ in the $n \to \infty$ limit 
\cite{wa23}.  We recall the definition that two graphs $G$ and $H$ are 
homeomorphic to each other, denoted as $G \sim H$, if one of them, say $H$, 
can be obtained from the other, $G$, by successive insertions of 
degree-2 vertices on bonds of $G$ \cite{biggsbook}.  
Each such insertion subdivides an existing edge of $G$ into two,
connected by the inserted degree-2 vertex.  This process is called
homeomorphic expansion and its inverse is called
homeomorphic reduction, i.e. the successive removal of vertices of degree 2
from a graph $H$.  Clearly, homeomorphic expansion of a graph always yields
another graph.  The inverse is not necessarily true; i.e., homeomorphic
reduction of a graph can produce a multigraph or pseudograph instead of a
(proper) graph \cite{graphdef}.  Here, a multigraph is a finite set of
vertices and bonds that, like a graph, has no bonds that loop around from a
given vertex back to itself but, in contrast to a (proper) graph,
may have more than one bond connecting two vertices.  A pseudograph is a finite
set of vertices and bonds that may have multiple bonds connecting two
vertices and may also have looping bonds.  For example, consider homeomorphic
reduction of a circuit graph $C_r$: removing one of the vertices (all of
which have degree 2), one goes from $C_r$ to $C_{r-1}$, and so forth, until one
gets to $C_3$.  During this sequence of homeomorphic reductions, one remains
within the category of graphs.  However, the next homeomorphic reduction takes
$C_3$ to $C_2$, which is a multigraph, not a proper graph.  Removing one of the
two vertices in $C_2$ produces the pseudograph $C_1$ 
consisting of a single vertex and
a bond that goes out and loops back to this vertex.  Thus homeomorphism is an
equivalence relation on pseudographs.  This complication will not
be relevant for our actual calculations of chromatic polynomials here, 
because we shall only use homeomorphic expansions, not
reductions, of graphs, and the homeomorphic expansion of a proper graph always
yields another proper graph. 

To investigate the
effects of homeomorphic expansions, it is natural to consider first the
simplest case, viz., an open (planar) 
chain of $p$-gons, i.e., polygons with $p$ 
vertices.  We shall picture the chain as having its longitudinal direction
oriented horizontally and its transverse direction oriented vertically, and
being comprised of $m$ $p$-gons, and thus $m+1$ vertical edges (see Fig. 1,
which will be discussed further below). 

\vspace{1 cm}
\begin{figure}
\centering
\leavevmode
\epsfxsize=4.0in
\epsffile{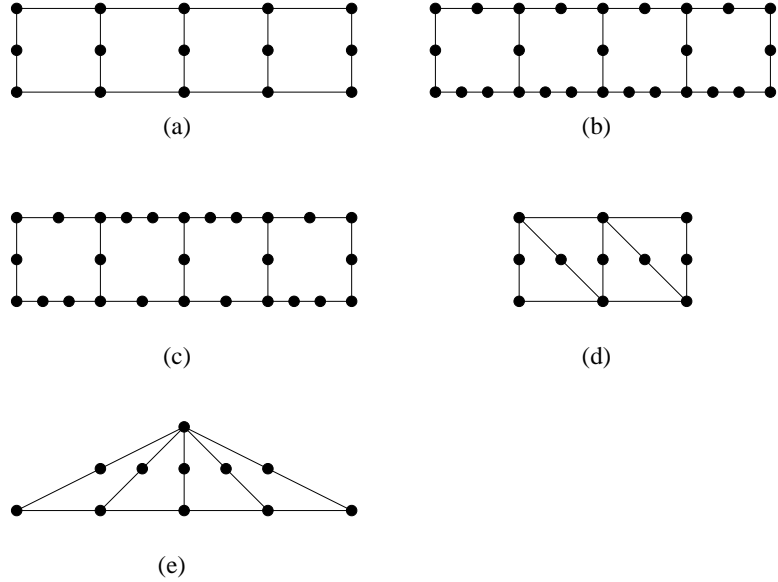}
\vspace{1 cm}
\caption{\footnotesize{
Illustration of homeomorphic expansions of open chains of $p$-gons.
(a) $(Ch)_{k_1,k_2,k,m}$ with $k_1=k_2=2$, whence $p=4$, and $k=3$, $m=4$; 
(b) $(Ch)_{k_1,k_2,\Sigma,k,m}$ with $k_1=3$, $k_2=4$, whence $p=7$, and 
$\Sigma=\Sigma_{+,m=4}=(+,+,+,+)$, $k=3$, $m=4$; 
(c) same as (b), but with $\Sigma=(+,-,-,+)$;
(d) $(Ch)_{k_1,k_2,\Sigma,k,m}$ with $k_1=1$, $k_2=2$, whence $p=3$, and 
$\Sigma=\Sigma_{alt.,m=4}=(+,-,+,-)$, $k=3$, $m=4$; 
(e) same as (d), but with $\Sigma=\Sigma_{+,m=4}$.}}
\label{hesgraph}
\end{figure}

The
property that the chain is open is equivalently stated as the property that it
has open or free boundary conditions in the longitudinal direction. 
Extending the notation used in \cite{wa,strip}, we consider an 
open chain of $m$ $p$-gons constructed such that a given $p$-gon intersects 
with the next $p$-gon in the chain along one of their mutual vertical edges, 
and such that for the $j$'th $p$-gon, there are $k_1$ vertices along either the
upper or the lower side from the vertex of a transverse (vertical) edge to 
the vertex of the next transverse edge (including these two latter vertices) 
and (ii) $k_2$ vertices along the other (=lower or upper, respectively) 
longitudinal side, 
from the vertex of the given transverse edge to the vertex of the next 
transverse edge, including these two latter vertices.  If $k_1 \ne k_2$, then 
for each of the $p$-gons in the chain, one has a choice as to whether to 
assign the $k_1$ vertices to the upper longitudinal side and $k_2$ vertices 
to the lower side, or vice versa.  To incorporate this
information in the $k_1 \ne k_2$ case, we define a parity vector 
\beq
\Sigma=(\sigma_1,\sigma_2,...\sigma_m)
\label{sigmavector}
\eeq
in which the $\sigma_j=+$ if for the
$j$'th $p$-gon, the assignment is (upper,lower) $=(k_1,k_2)$  and
$\sigma_j=-$ if the assignment is (upper,lower) $=(k_2,k_1)$.
We shall denote this strip graph as 
\beq
(Ch)_{k_1,k_2,\Sigma,m}
\label{chk1k2sm}
\eeq
with the understanding that if $k_1=k_2$, $\Sigma$ is not defined and can be
omitted in the notation:
\beq
(Ch)_{k_1,k_1,m}
\label{chk1k1m}
\eeq
Thus, $(Ch)_{2,2,m}$ represents the strip (chain) of $m$ squares, with each 
successive pair of squares intersecting on a common edge. 
It will be convenient to define two special $\Sigma$ vectors: 
\beq
\Sigma_{+,m} = (+,+,...,+)
\label{sigmaplus}
\eeq
and, for the case of alternating signs,
\beq
\Sigma_{alt.,m} = (+,-,+,-,...)
\label{sigmaalt}
\eeq
where the dimension $m$ of $\Sigma$ is indicated explicitly here but will
sometimes be implicit below. Obviously, the $(Ch)_{k_1,k_2,\Sigma,m}$ graphs 
are invariant under a reflection about the 
longitudinal axis, which in the $k_1 \ne k_2$ case 
amounts to a simultaneous reversal of the signs of all
of the $\sigma_j$'s.  Let us define this parity operation on $\Sigma$ as 
\beq
P(\Sigma) = -\Sigma
\label{psigma}
\eeq
Then 
\beq
(Ch)_{k_1,k_2,\Sigma,m}= (Ch)_{k_1,k_2,P(\Sigma),m} = (Ch)_{k_2,k_1,\Sigma,m}
\label{ch1221}
\eeq
The total number of vertices in each $p$-gon is 
\beq
p = k_1 + k_2
\label{pvertices}
\eeq
and the total number of vertices in the chain is 
\beq
n = v(Ch_{k_1,k_2,\Sigma,m}) = (p-2)m+2
\label{chmn}
\eeq
The chromatic polynomial for this open chain $(Ch)_{k_1,k_2,\Sigma,m}$ is 
\beq
P((Ch)_{k_1,k_2,\Sigma,m},q) = q(q-1)D_p(q)^m
\label{pchainmpgons}
\eeq
Observe the important properties that $n$ and the chromatic polynomial 
$P((Ch)_{k_1,k_2,\Sigma,m},q)$ 
(i) depend on $k_1$ and $k_2$ only through their sum, 
$p=k_1+k_2$; and (ii) are independent of $\Sigma$.  These properties were 
implicit in 
the compact notation that we used previously in Ref. \cite{wa}, viz., simply
$(Ch)_{p,m}$, representing any of the specific $p$-gon chain graphs 
$(Ch)_{k_1,k_2,\Sigma,m}$ such that $k_1+k_2$ satisfies the condition 
(\ref{pvertices}).  This provides an example of the fact that there
is not a 1--1 correspondence between graphs and chromatic polynomials of
graphs; different graphs may have the same chromatic polynomial.  In the 
present case, all of the different graphs $(Ch)_{k1,k2,\Sigma,m}$ satisfying 
the condition (\ref{pvertices}) have the same chromatic polynomial, eq. 
(\ref{pchainmpgons}).  If $k_1 \ne k_2$, then, taking account of the overall 
symmetry (\ref{ch1221}), this number is given by $2^{m-1}$.  Of course, if 
$k_1=k_2$, then there is only one such graph. 

For $k_1 \ge 2$ and $k_2 \ge 2$, we may view the family 
$P((Ch)_{k_1,k_2,\Sigma,m},q)$ as being constructed by means of homeomorphic
expansion of the strip of $m$ squares, $(Ch)_{2,2,m}$: one performs 
homeomorphic expansions by inserting $k_1-2$ degree--2 vertices on the upper or
lower edge of the first square and $k_2-2$ degree--2 vertices on the other
side (lower or upper, respectively) of this first square, making another
homeomorphic expansion on the second square, and so forth along the strip. 
The resultant homeomorphically expanded strip is
precisely $(Ch)_{k_1,k_2,\Sigma,m}$; 
that is, denoting this homeomorphic expansion 
(HE) of the \underline longitudinal edges as $HEL_{k_1-2,k_2-2,\Sigma}$, 
we have 
\beq
HEL_{k_1-2,k_2-2,\Sigma}((Ch)_{2,2,m}) = (Ch)_{k_1,k_2,\Sigma,m}
\label{helrel}
\eeq
In the
$L_m$ limit of eq. (\ref{minf}), i.e., $m \to \infty$ with fixed $k_1$ and
$k_2$, the resultant nonanalytic locus ${\cal B} = \emptyset$, i.e., $W$ is
analytic in the full complex $q$ plane, and is given by 
\beq
W([\lim_{m \to \infty}(Ch)_{k_1,k_2,\Sigma,m}],q) = (D_p)^{1/(p-2)} 
\quad {\rm where} \quad p=k_1+k_2
\label{whel}
\eeq

We next proceed to consider the more complicated case of homeomorphic 
expansions of the transverse edges on the strip of $p$-gons, 
including those shared by adjacent $p$-gons 
and the two edges on the transverse ends of the strip of 
squares. We let $k$ denote
the number of vertices on each of these $m+1$ transverse edges, including the
pair of vertices belonging to the original strip before the
homeomorphic expansion.   The resultant family of graphs obtained by this
process of homeomorphic expansion of \underline transverse edges is 
\beq
HET_{k-2}((Ch)_{k_1,k_2,\Sigma,m}) \equiv (Ch)_{k_1,k_2,\Sigma,k,m}
\label{het}
\eeq
An illustration is given in Fig. \ref{hesgraph}(a) for the case $m=4$ and 
$k=3$. 
For $k_1 \ge 2$ and $k_2 \ge 2$, one may equivalently regard the family 
$(Ch)_{k_1,k_2,\Sigma,k,m}$ as being 
constructed by the simultaneous combined homeomorphic expansions of the
longitudinal and transverse edges of a strip of $m$ squares:
\beq
(Ch)_{k_1,k_2,\Sigma,k,m} = HET_{k-2}\biggl [ HEL_{k_1-2,k_2-2,\Sigma}
((Ch)_{2,2,m})\biggr ]
\label{helt}
\eeq
Again, the graph is the same if one reflects it about the longitudinal axis:
\beq
(Ch)_{k_1,k_2,\Sigma,k,m} = (Ch)_{k_1,k_2,P(\Sigma),k,m} = 
(Ch)_{k_2,k_1,\Sigma,k,m}
\label{chk1k2kmsym}
\eeq
One illustration is given in Fig. \ref{hesgraph}(b) for the case $m=4$, $k=3$, 
$k_1=3$, and $k_2=4$, whence $p=7$, and $\Sigma=\Sigma_{+,m=4}=(+,+,+,+)$.  
A second 
illustration is Fig. \ref{hesgraph}(c) for the same values of $m$, $k$, $k_1$,
and $k_2$ but $\Sigma=(+,-,-,+)$.  
One can also relate certain of these
homeomorphic expansions to strip graphs studied in Ref. \cite{strip}: 
\beq
(Ch)_{2,2,k,m} = HET_{k-2}(G_{sq(L_y=2),m})
\label{hetsq}
\eeq
\beq
(Ch)_{1,2,\Sigma_{alt.},k,m} = HET_{k-2}(G_{t(L_y=2),m})
\label{hettri}
\eeq
and
\beq
(Ch)_{3,3,k,m} = HET_{k-2}(G_{hc(L_y=2),m})
\label{hethc}
\eeq
Here, the abbreviations $sq$, $t$, and $hc$ stand for strips of the square,
triangular, and honeycomb lattice, and $L_y$ denotes the number of vertices in
the vertical direction. 

\begin{figure}
\centering
\leavevmode
\epsfxsize=4.0in
\epsffile{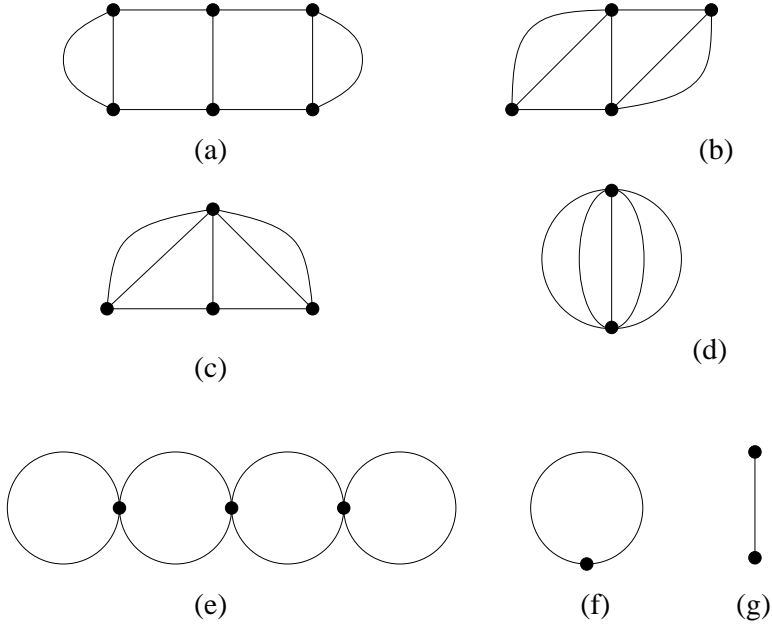}
\vspace{1 cm}
\caption{\footnotesize{Illustrations of homeomorphic classes for 
$(Ch)_{k_1,k_2,\Sigma,k,m}$ with $m=4$: 
(a) $k_1 \ge 2$, $k_2 \ge 2$ (whence $p \ge 4$), eq. (\ref{squarehomclass});
(b) $k_1=1$, $k_2 \ge 2$ (whence $p \ge 3$), 
$\Sigma=\Sigma_{alt.}$, eq. (\ref{trihomclass});
(c) $k_1=1$, $k_2 \ge 2$ (whence $p \ge 3$), $\Sigma=\Sigma_+$, one member of 
the set of classes in eq. (\ref{trimhomclass}); 
(d) $k_1=1$, $k_2=1$, so $p=2$, eq. (\ref{hkrhomclass}); 
(e) $p \ge 4$, $k=1$, eq. (\ref{circuithomclass}); 
(f) $m=1$, eq. (\ref{m1homclass}); 
(g) $m=0$, eq. (\ref{m0homclass}).}}
\label{homclass}
\end{figure}

It is also possible for $p$ to be less than
4, in which case the graph does not represent the homeomorphic expansion of an
original strip of squares.  Two illustrations of this are given in Fig. 
\ref{hesgraph}, both with $p=3$.  With no loss of generality, we take 
$k_1=1$ for these examples.  Figs. \ref{hesgraph}(d,e) show graphs with 
$k_1=1$, $k_2=2$, $k=3$, $m=4$, and $\Sigma=\Sigma_{alt.,m=4}$ and 
$\Sigma_{+,m=4}$, respectively.  
Note that in the limit $m \to \infty$, the family 
$(Ch)_{1,2,\Sigma,k,m}$ with $\Sigma=(+,+,...,+)$ contains one vertex of 
infinite degree.  For $p=2$ and $k=2$, the chain is actually not a (proper)
graph but a multigraph (see Fig. \ref{homclass}(d)), which we will not
consider.  For $p=2$ and $k \ge 3$, the chain degenerates according to 
\beq
(Ch)_{p=2,k,m} \equiv (Ch)_{1,1,k,m} = 
HEC_{k-3}(\overline K_2 + \overline K_m) \equiv H_{k,m}
\label{p2hkm}
\eeq
which we have previously studied in detail \cite{wa23,complete} (where the 
notation $HEC$ was used to refer to the 
homeomorphic expansion of the bonds \underline connecting the 
vertices in the $\overline K_2$ subgraph to the vertices in the 
$\overline K_m$ subgraph).  

Two special cases are
\beq
(Ch)_{p,k,m=0} = T_k
\label{chpkm0}
\eeq
and
\beq
(Ch)_{p,k,m=1} = C_{p+2k-4}
\label{chpkm1}
\eeq
where $T_n$ is a tree graph \cite{complete} and $C_n$ is the circuit graph.

\vspace{8mm}

Thus, the families $(Ch)_{k_1,k_2,\Sigma,k,m}$ include the following
homeomorphic classes, as illustrated in Fig. \ref{homclass} (here $HR$
denotes homeomorphic reduction, and $m \ge 1$ in eqs. (\ref{squarehomclass}) to
(\ref{circuithomclass})): 

\vspace{6mm}

\beqs
(Ch)_{k_1,k_2,\Sigma,k,m} & \sim & (Ch)_{2,2,2,m} = G_{sq(L_y=2),m} 
\sim HR(G_{sq(L_y=2),m}) \cr
& & \quad {\rm for} \quad k_1 \ge 2 \ ,\ k_2 \ge 2 \ , \ 
{\rm and} \quad k \ge 2
\label{squarehomclass}
\eeqs
\beq
(Ch)_{1,k_2,\Sigma_{alt.},k,m} \sim (Ch)_{1,2,\Sigma_{alt.},2,m} =
G_{t(L_y=2),m} \sim HR(G_{t(L_y=2),m}) 
\quad {\rm for} \quad k_2 \ge 2 \quad {\rm and} \quad k \ge 2
\label{trihomclass}
\eeq
\beq
\biggl \{ 
(Ch)_{1,k_2,\Sigma,k,m} \sim HR((Ch)_{1,2,\Sigma,2,m}) \quad {\rm for} \quad 
k_2 \ge 2 \ , \ \ k \ge 2 \ , \quad {\rm and} \quad \Sigma \ne \Sigma_{alt.} 
\biggr \}
\label{trimhomclass}
\eeq
\beq
(Ch)_{1,1,k,m} \sim (Ch)_{1,1,2,m} \quad {\rm for} \quad k \ge 2
\label{hkrhomclass}
\eeq
\beq
(Ch)_{k_1,k_2,\Sigma,k=1,m} = \bigcap_{m;T_1} C_{p+2k-4} 
\sim \bigcap_{m;T_1} C_2 \sim C_1 \cdot \Bigl [\bigcap_{m-2;T_1} C_2
\Bigr ] \cdot C_1 \quad {\rm for} \quad p \ge 4
\label{circuithomclass}
\eeq
\beq
(Ch)_{k_1,k_2,\Sigma,k,m=1} \sim C_1
\label{m1homclass}
\eeq
\beq
(Ch)_{k_1,k_2,\Sigma,k,m=0} \sim T_2
\label{m0homclass}
\eeq

\vspace{6mm}

\noindent
The families $G_{sq(L_y),m}$ and $G_{t(L_y),m}$ were defined above, and 
some relevant graph theory definitions are given in \cite{complete}. 
The notation $\{...\}$ in eq. (\ref{trimhomclass}) indicates that this set 
actually subsumes multiple homeomorphism classes.   In 
eq. (\ref{circuithomclass}), the notation $\bigcap_{m;T_1} G$ means  
the chain of $m$ repeated
graphs $G$, each of which intersects the next in a single vertex ($=T_1$), and
the notation $C_1 \cdot G$ means the chain in which the $C_1$ pseudograph
intersects with the rest of the chain, $G$, at its single vertex.  
In our introductory discussion about homeomorphic classes we mentioned how, in
general, homeomorphic reductions of graphs yield objects that are not
(proper) graphs,
but instead are multigraphs or pseudographs.  Specifically, here, 
$HR(G_{sq(L_y=2),m})$ in eq. (\ref{squarehomclass}), illustrated in 
Fig. \ref{homclass}(a), is a multigraph, as are $HR(G_{t(L_y=2),m})$ in eq. 
(\ref{trihomclass}) (Fig. \ref{homclass}(b)), 
$HR((Ch)_{1,2,\Sigma,2,m})$ in eq. 
(\ref{trimhomclass}) (Fig. \ref{homclass}(c)), and 
$(Ch)_{1,1,2,m}$ in eq. (\ref{hkrhomclass}) 
(Fig. \ref{homclass}(d)), while 
$ C_1 \cdot \Bigl [\bigcap_{m-2;T_1} C_2 \Bigr ] \cdot C_1$ in eq. 
(\ref{circuithomclass}) (Fig. \ref{homclass}(e)) and $C_1$ in 
eq. (\ref{m1homclass}) (Fig. \ref{homclass}(f)) are pseudographs.  In contrast,
the homeomorphic reduction $(Ch)_{k_1,k_2,\Sigma,k,m=0} \sim T_2$ in eq. 
(\ref{m0homclass}) (Fig. \ref{homclass}(g)) does yield a proper graph. 

The number of vertices of $(Ch)_{k_1,k_2,\Sigma,k,m}$ is
\beq
n=v((Ch)_{k_1,k_2,\Sigma,k,m}) = (k+p-4)m+k
\label{verticesch}
\eeq
and the girth (= length of minimum length circuit) is
\beq
g((Ch)_{k_1,k_2,\Sigma,k,m}) = 2k+p-4
\label{girthch}
\eeq
where in both cases, $p=k_1+k_2$ as given by eq. (\ref{pvertices}).  From
eq. (\ref{girthch}), it follows that for the nontrivial case 
$m \ge 1$, the chromatic number 
\cite{chromatic} is (independently of $k$)
\beq
\chi((Ch)_{k_1,k_2,\Sigma,k,m}) = \cases{ 2 & if $p$ is even \cr
                                   3 & if $p$ is odd \cr }
\label{chi}
\eeq
($\chi=2$ for the trivial case $m=0$.) Hence, if $m \ge 1$, 
\beq
P((Ch)_{k_1,k_2,\Sigma,k,m},q=2) = 0 \quad {\rm if} \quad p \quad 
{\rm is \ odd} 
\label{pq2zeropodd}
\eeq
As well as its obvious importance for the chromatic polynomial, the value of
$\chi$ also affects the properties of the asymptotic function $W$, as will 
be evident in our explicit results below.  We also recall our notation 
$q_c$ for the maximal finite real point at which $W$ is nonanalytic 
\cite{p3afhc,w}.  

\section{Generating Function, Chromatic Polynomials, and $W$ Function for 
Homeomorphic Families of Strip Graphs}

We now calculate a generating function for the chromatic polynomial of the 
$(Ch)_{k1,k2,\Sigma,k,m}$ strip graph. Our method 
uses the deletion-contraction theorem to derive a set of recursion relations
involving $(Ch)_{k_1,k_2,\Sigma,k,m}$ and certain other ancillary families of
graphs, as in Ref. \cite{strip}.  We shall initially denote the generating 
function as $\Gamma((Ch)_{k_1,k_2,\Sigma,k},q,x)$, where $x$ is an auxiliary 
expansion variable, and yields the chromatic polynomial 
$P((Ch)_{k_1,k_2,\Sigma,k,m},q)$ as the coefficient of $x^m$ in the Taylor 
series expansion of the generating function about $x=0$: 
\beq
\Gamma((Ch)_{k_1,k_2,\Sigma,k},q,x)=
\sum_{m=0}^{\infty}x^m P((Ch)_{k_1,k_2,\Sigma,k,m},q),
\label{chgfdef}
\eeq
The $m=0$ term is just the chromatic polynomial of the end-rung of the chain, 
\beq
P(Ch_{k_1,k_2,\Sigma,k,m=0},q) = P(T_k,q) = q(q-1)^{k-1}
\label{chgfm0}
\eeq
As before \cite{strip}, we find that the generating function is a rational
function in the auxiliary variable $x$ and, separately, in the variable $q$.
In particular, we find that the degrees of the numerator and denominator, as
polynomials in $x$, are 
\beq
d_{\cal N}=1
\label{degxn}
\eeq
\beq
d_{\cal D}=2
\label{degxd}
\eeq
i.e., 
\beq
\Gamma((Ch)_{k_1,k_2,\Sigma,k},q,x)=
\frac{{\cal N}((Ch)_{k_1,k_2,\Sigma,k},q,x)}
{{\cal D}((Ch)_{k_1,k_2,\Sigma,k},q,x)}=\frac{A_0+A_1 x}{1+ b_1 x + b_2 x^2} 
\label{chgf}
\eeq
Two important properties are that the generating function and chromatic 
polynomial for the homeomorphically expanded graph 
$(Ch)_{k_1,k_2,\Sigma,k,m}$ (i) are independent of $\Sigma$ and 
(ii) depend on $k_1$ and $k_2$ only through their sum, as given in eq. 
(\ref{pvertices}):
\beq
\{ \Gamma((Ch)_{k_1,k_2,\Sigma,k},q,x), \quad P((Ch)_{k_1,k_2,\Sigma,k,m},q) 
\} = fns.(p,k,m) \quad {\rm where} \quad p=k_1+k_2
\label{pdependence}
\eeq
Henceforth we shall incorporate this property in our notation, as follows: we
shall use the symbol $(Ch)_{p,k,m}$ to encompass the full class of families 
$(Ch)_{k_1,k_2,\Sigma,k,m}$ such that $k_1+k_2=p$: 
\beq
(Ch)_{p,k,m} \equiv \{ (Ch)_{k_1,k_2,\Sigma,k,m} \}
\label{chpkm}
\eeq
and similarly, 
\beq
P((Ch)_{p,k,m},q) \equiv P((Ch)_{k_1,k_2,\Sigma,k,m},q)
\label{pchpkm}
\eeq
and 
\beq
\Gamma((Ch)_{p,k},q,x) \equiv \Gamma((Ch)_{k_1,k_2,\Sigma,k},q,x) 
\label{gammapk}
\eeq
Since, as noted above in eq. (\ref{p2hkm}), for $p=2$, the chain reduces to 
$HEC_{k-3}(\overline K_2 + \overline K_m)$, which we have already studied in 
detail in Ref. \cite{wa23}, we restrict ourselves to the cases $p \ge 3$ here. 
Since for $k=2$, the elementary result 
(\ref{pchainmpgons}) holds, we shall concentrate on the case $k \ge 3$. 

We calculate 
\beq
A_0=P((Ch)_{p,k,m=0},q) = q(q-1)^{k-1}
\label{a0ch}
\eeq
(independent of $p$) and 
\beq
A_1=-q^2(q-1)(-1)^pD_k D_{k-1}
\label{a1ch}
\eeq
\beq
b_1=-(-1)^p D_k-(q-1)D_{k+p-3}
\label{b1ch}
\eeq
\beq
b_2=(-1)^p(q-1)^{p-1} D_kD_{k-1}
\label{b2ch}
\eeq
Factorizing the denominator of the generating function as in
eq. (\ref{lambdaform}) gives 
\beq
{\cal D}((Ch)_{p,k},q,x)=(1-\lambda_1x)(1-\lambda_2x)
\label{dfactors}
\eeq
where $\lambda_{1,2}=-(1/2)[b_1 \pm \sqrt{R}]$ with the discriminant 
$R=b_1^2-4b_2$, i.e.,
\beq
\lambda_{1,2}=\frac{1}{2} \biggl [ (-1)^pD_k+(q-1)D_{k+p-3} \pm 
\sqrt{R} \biggr ],
\label{lambda12}
\eeq
where
\beqs
R &=& (D_k)^2+[(q-1)D_{k+p-3}]^2+2(-1)^p(q-1)D_k D_{k+p-3} \cr
& & - 4 (-1)^p(q-1)^{p-1} D_k D_{k-1}
\label{disc}
\eeqs
The degree of $R$, as a polynomial in $q$, is
\beq
deg_q(R) = 2(k+p-4)
\label{degqr}
\eeq
and the term containing the highest power of $q$ in $R$ is 
$[(q-1)D_{k+p-3}]^2$.  The chromatic polynomial is thus given by the general
formulas, eq. (\ref{pgsumkmax2}) or (\ref{chrompgsumlam})--(\ref{tjkappa1}) 
with eqs. (\ref{degxn}), (\ref{degxd}), 
\beq
c_1 = \Biggl [ \frac{A_1 + A_0 \lambda_1}
{\lambda_1-\lambda_2} \Biggr ] (\lambda_1)^{-k/(k+p-4)}
\label{c1}
\eeq
\beq
c_2 = \Biggl [ \frac{A_1 + A_0 \lambda_2}
{\lambda_2-\lambda_1} \Biggr ] (\lambda_2)^{-k/(k+p-4)}
\label{c2}
\eeq
$a_j=\lambda_j$, $j=1,2$, and 
\beq
t_j = \frac{1}{\kappa_1} = \frac{1}{k+p-4} \quad {\rm for} \quad j=1,2
\label{kappa1}
\eeq
In the $m \to \infty$ limit, 
\beq
W([\lim_{m \to \infty}(Ch)_{p,k,m}],q) = (\lambda_{max})^{1/(k+p-4)}
\label{wlambda}
\eeq
where $\lambda_{max}$ refers to the $\lambda_j$ ($j=1$ or 2) in eq. 
(\ref{lambda12}) with maximal absolute magnitude.  The continuous locus of
nonanalyticities ${\cal B}$ in $W([\lim_{m \to \infty}(Ch)_{p,k,m}],q)$ is
given by the locus of solutions of the equation 
\beq
|\lambda_1|=|\lambda_2|
\label{degeneq}
\eeq
as in the general eq. (\ref{lambdamax}). 

\section{Explicit Examples}

Our exact calculations of the respective continuous nonanalytic loci 
${\cal B}$ are shown for $p=3$ and $p=4$, and for each $p$, 
several values of $k \ge 3$, in 
Figs. \ref{hep3k34}--\ref{hep4k78}.  Further results for $p=5$ and $p=6$ are
given in Appendix 2. 
(As we have discussed earlier, with the ordering of the limits in eq. 
(2.13) of Ref. \cite{w}, $W$ has no discrete isolated nonanalytic points.)
Because a chromatic polynomial 
$P(G,q)$ has real (indeed, integer) coefficients, ${\cal B}$ has the basic 
property of remaining invariant under the replacement $q \to q^*$:
${\cal B}(q) = {\cal B}(q^*)$. 
For lack of space we have only included some of our results; plots for other 
values of $k$ up to $k=10$ for various values of $p$ 
are given in Ref. \cite{thesis}.  As the number of
vertices of a graph $n \to \infty$, the nonanalytic locus ${\cal B}$ forms 
by the coalescence of a subset of the zeros of the chromatic polynomials 
(= chromatic zeros) of the graph. Accordingly, 
it is also of interest to plot these chromatic zeros for reasonably long  
finite--length chains in the figures, to assess how closely these approach 
the asymptotic curves and line segments.  The reason that slightly smaller
values of $m$ are used for families with larger values of $k$ and/or $p$ is
that, owing to the relation (\ref{verticesch}), these lead to comparably large
number of total vertices. 

\pagebreak

\begin{figure}
\vspace{-4cm}
\centering
\leavevmode
\epsfxsize=3.0in
\begin{center}
\leavevmode
\epsffile{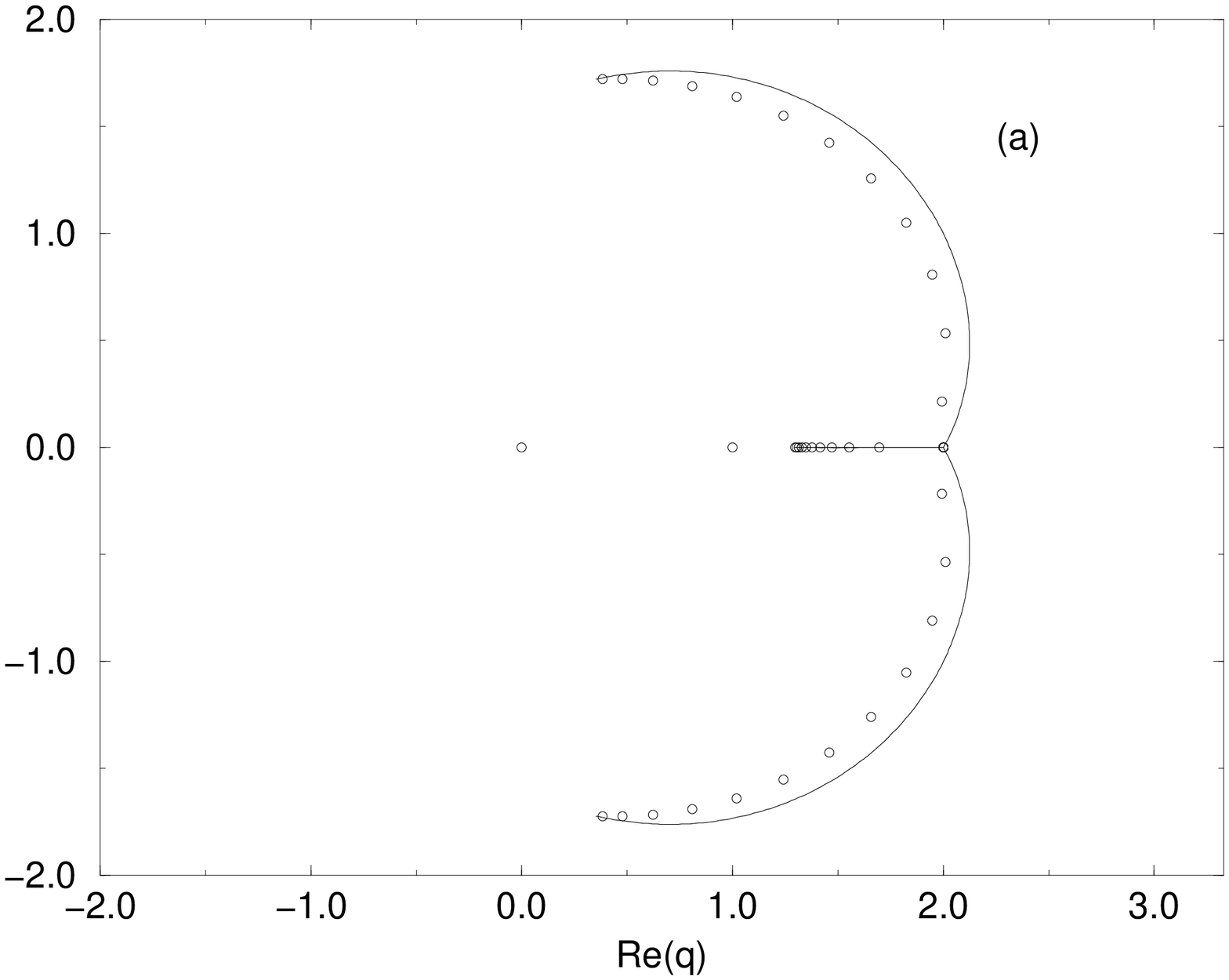}
\end{center}
\vspace{-3cm}
\begin{center}
\leavevmode
\epsfxsize=3.0in
\epsffile{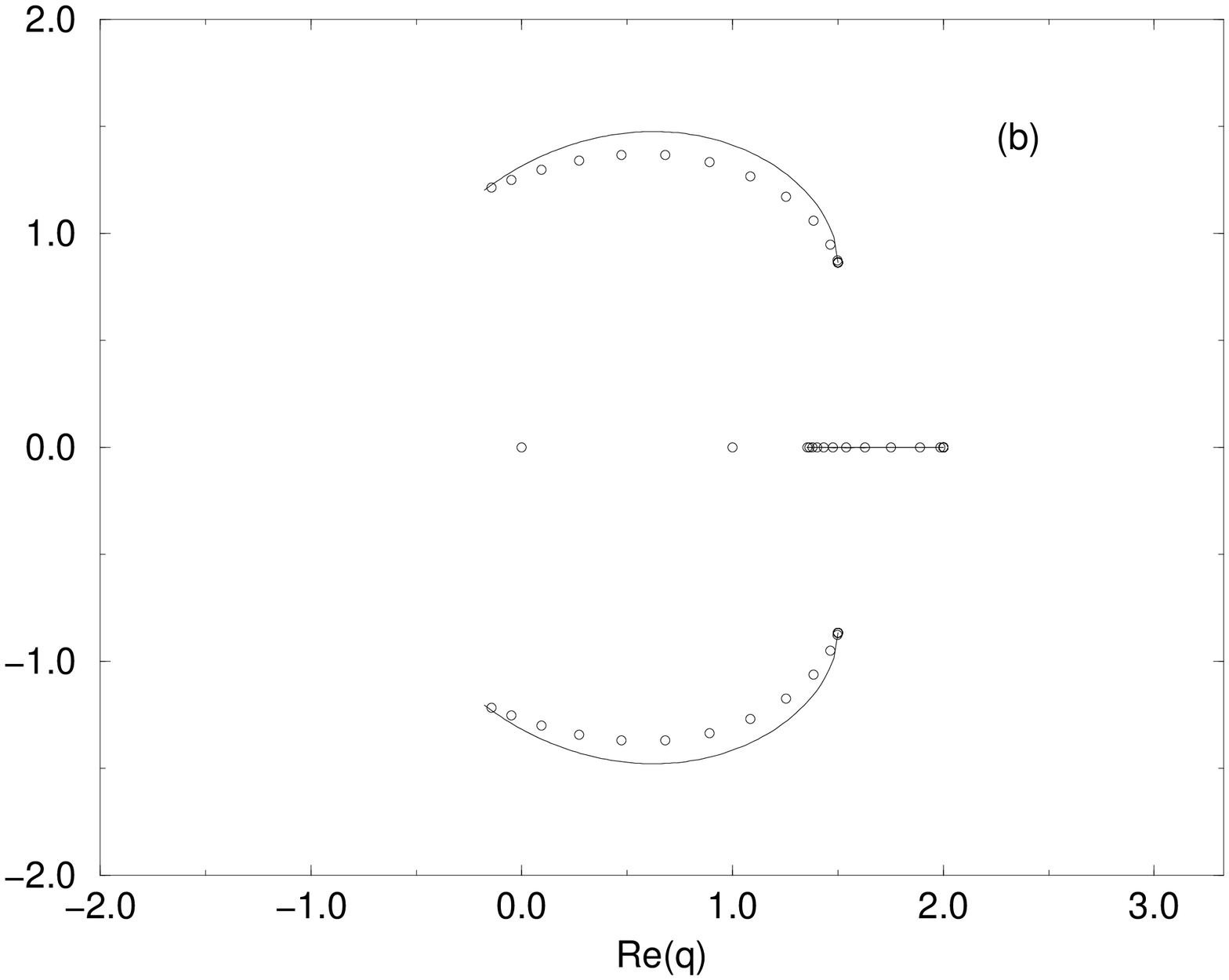}
\end{center}
\vspace{-2cm}
\caption{\footnotesize{Locus ${\cal B}$ and analytic structure of the 
function $W([\lim_{m \to \infty}(Ch)_{p,k,m}],q)$ for $p=3$ and $k=$ (a) 3 (b)
4.  For comparison, chromatic zeros are shown for $m=22$.}}
\label{hep3k34}
\end{figure}

\pagebreak 

\begin{figure}
\vspace{-4cm}
\centering
\leavevmode
\epsfxsize=3.0in
\begin{center}
\leavevmode
\epsffile{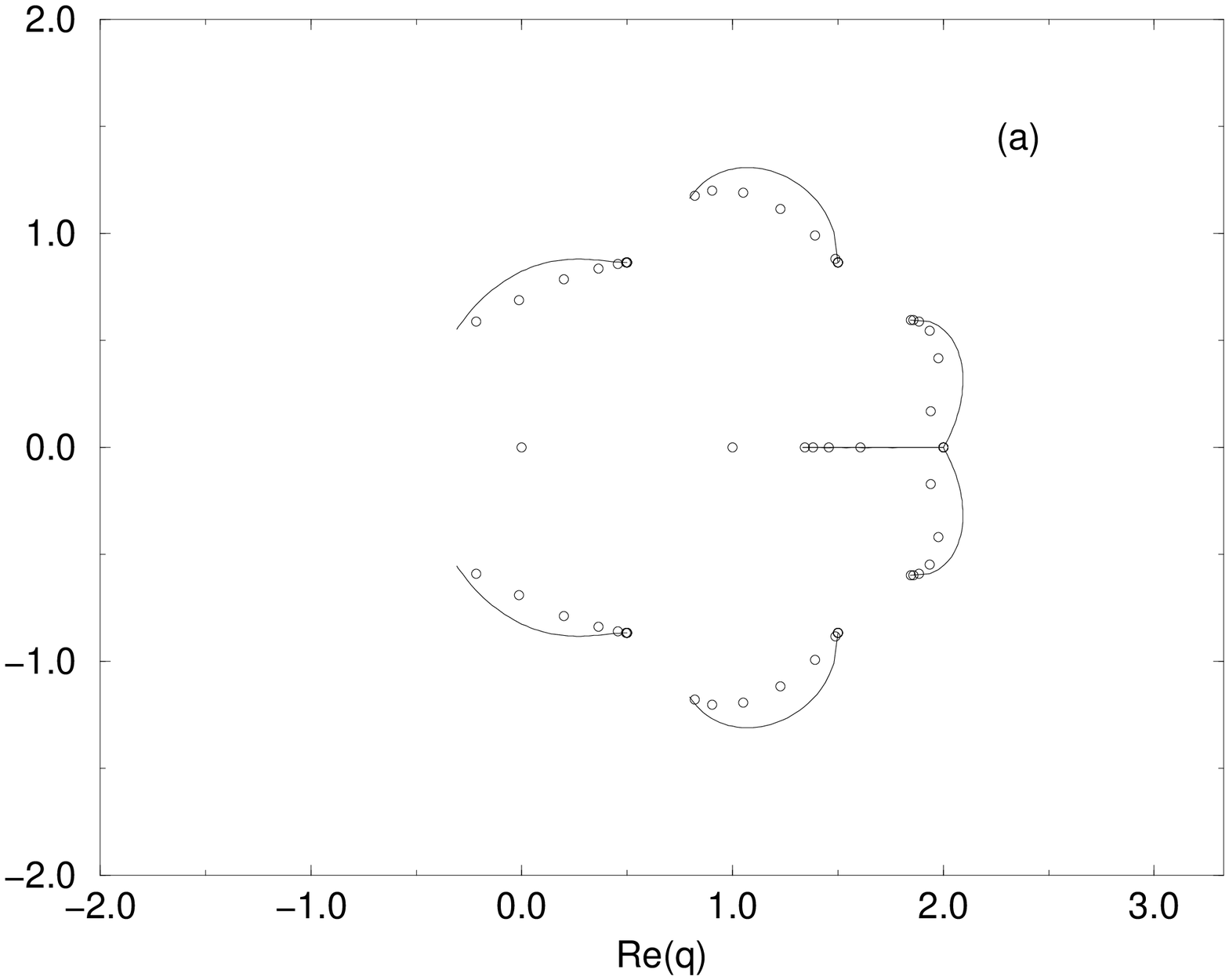}
\end{center}
\vspace{-3cm}
\begin{center}
\leavevmode
\epsfxsize=3.0in
\epsffile{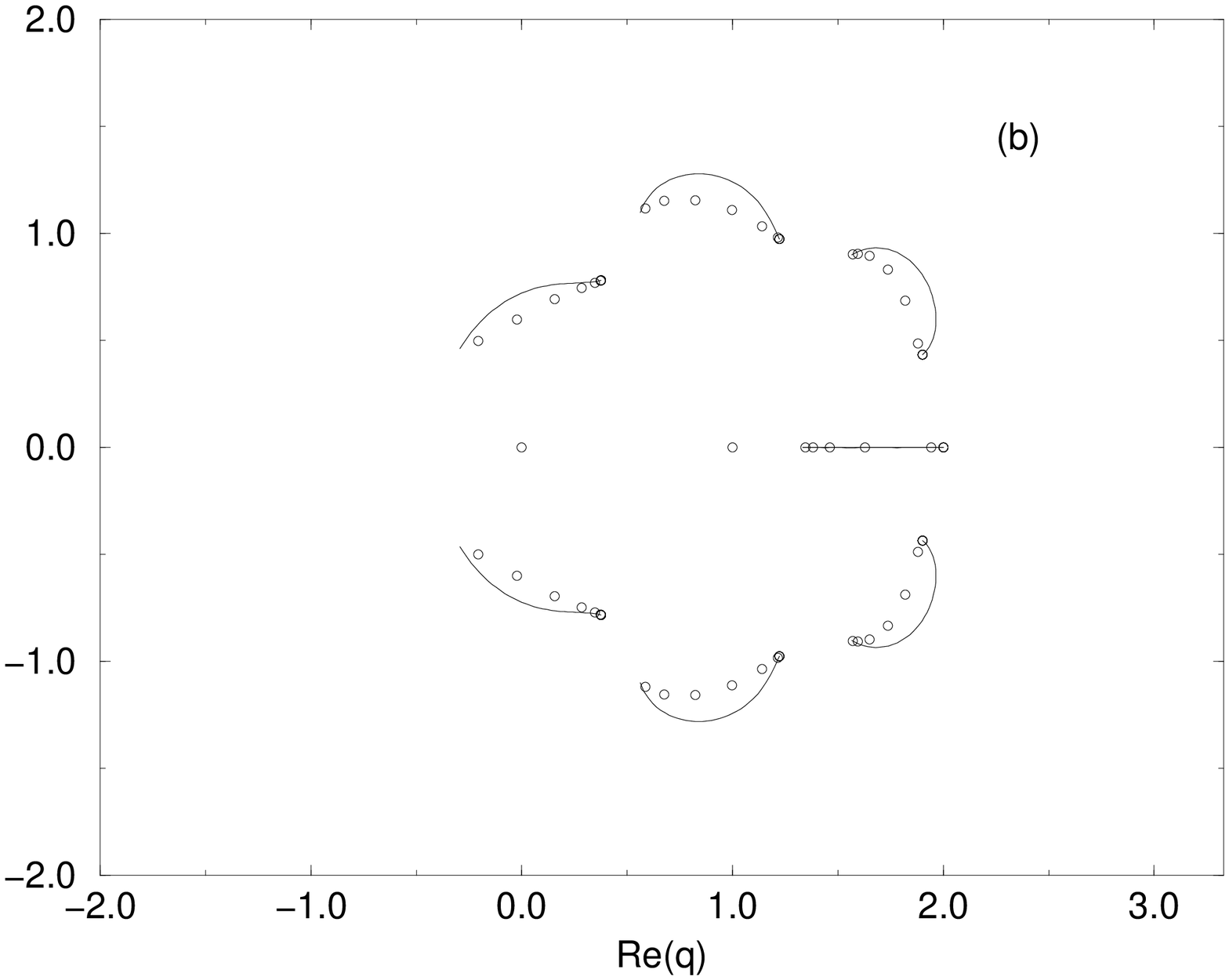}
\end{center}
\vspace{-2cm}
\caption{\footnotesize{As in Fig. \ref{hep3k34} for $p=3$ and $k=$ (a) 7 (b)
8.  For comparison, chromatic zeros are shown for $m=10$.}}
\label{hep3k78}
\end{figure}

\pagebreak 

\begin{figure}
\vspace{-4cm}
\centering
\leavevmode
\epsfxsize=3.0in
\begin{center}
\leavevmode
\epsffile{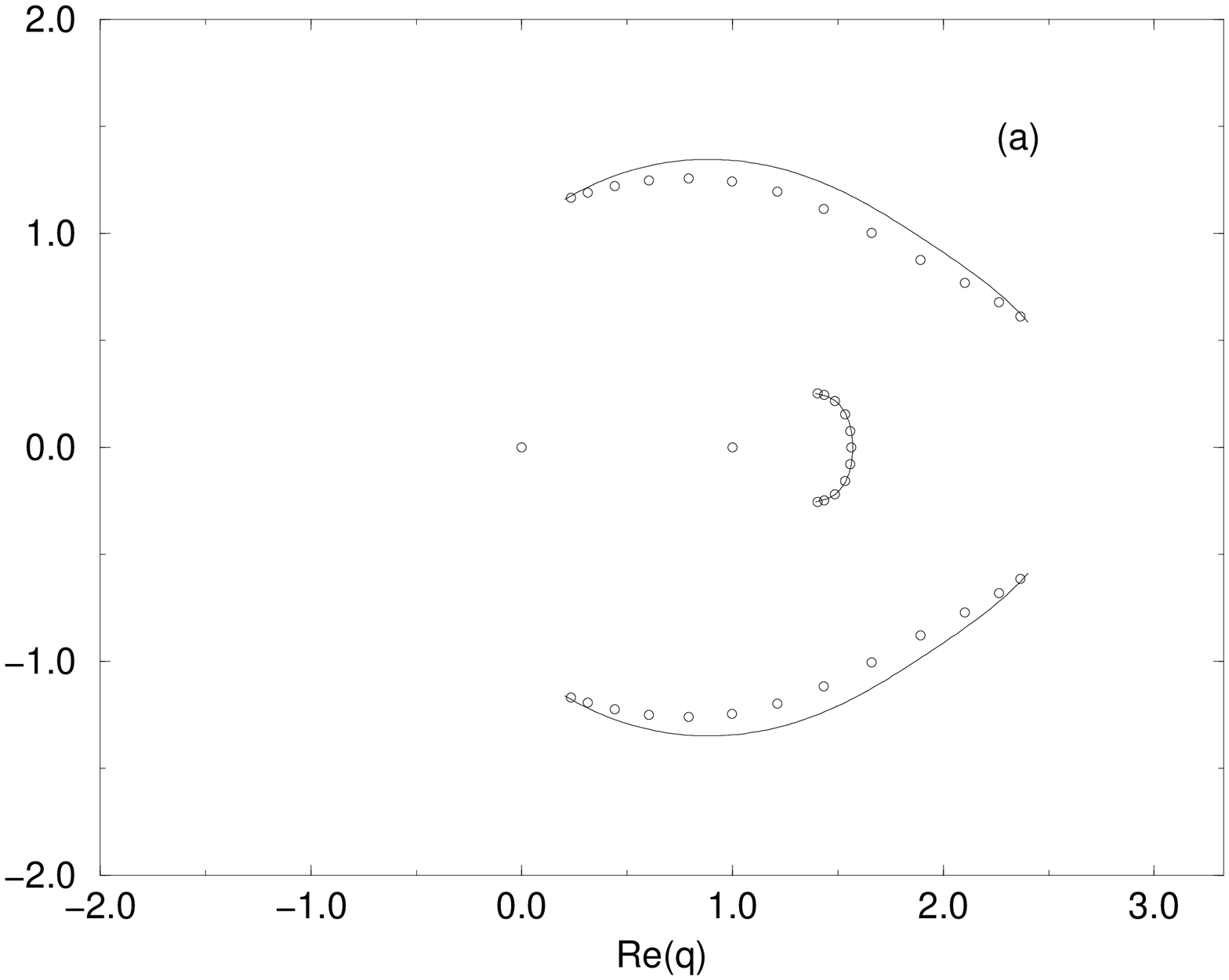}
\end{center}
\vspace{-3cm}
\begin{center}
\leavevmode
\epsfxsize=3.0in
\epsffile{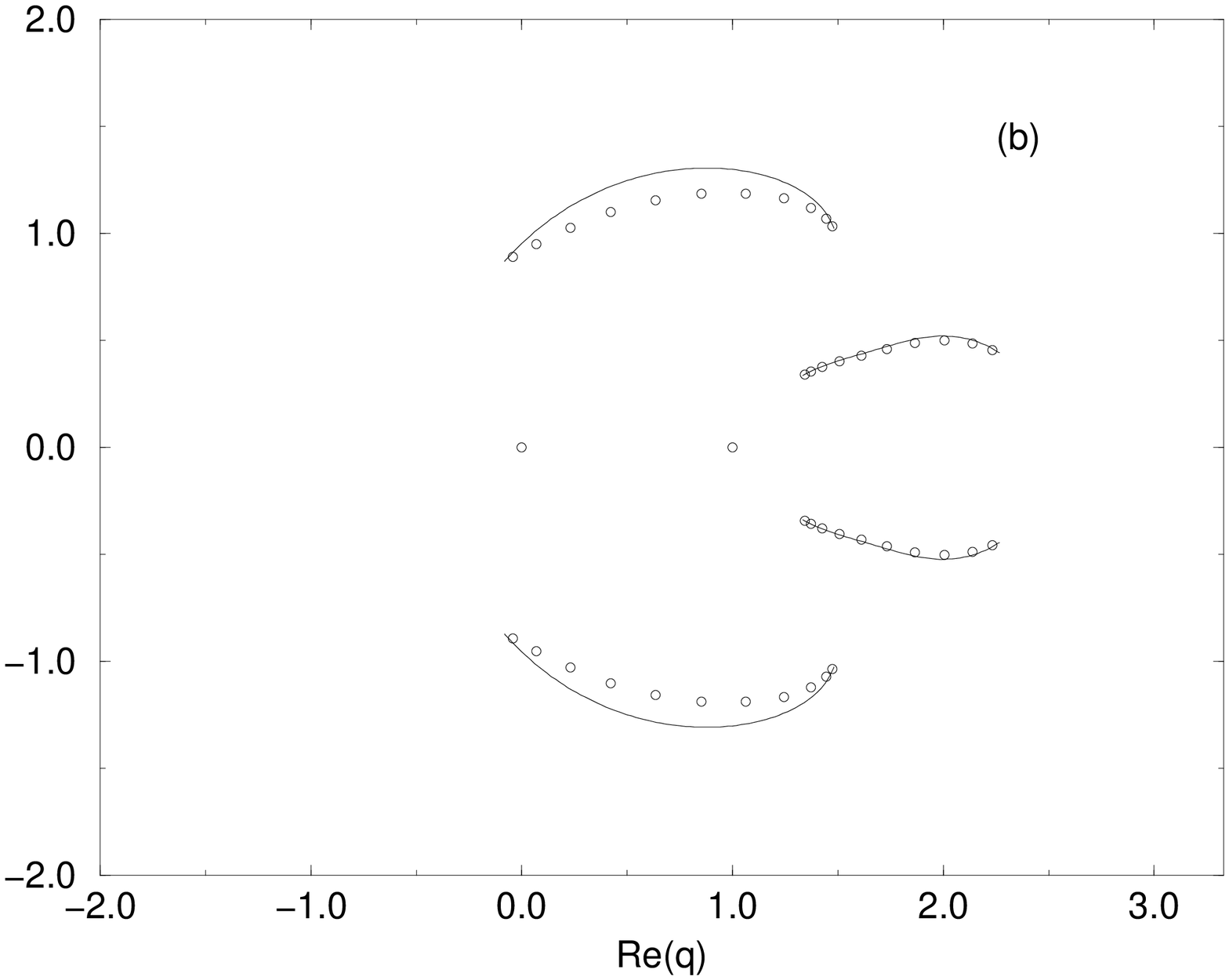}
\end{center}
\vspace{-2cm}
\caption{\footnotesize{As in Fig. \ref{hep3k34} for $p=4$ and $k=$ (a) 3 (b)
4.  For comparison, chromatic zeros are shown for $m=$ (a) 12 (b) 10.}}
\label{hep4k34}
\end{figure}

\pagebreak

\begin{figure}
\vspace{-4cm}
\centering
\leavevmode
\epsfxsize=3.0in
\begin{center}
\leavevmode
\epsffile{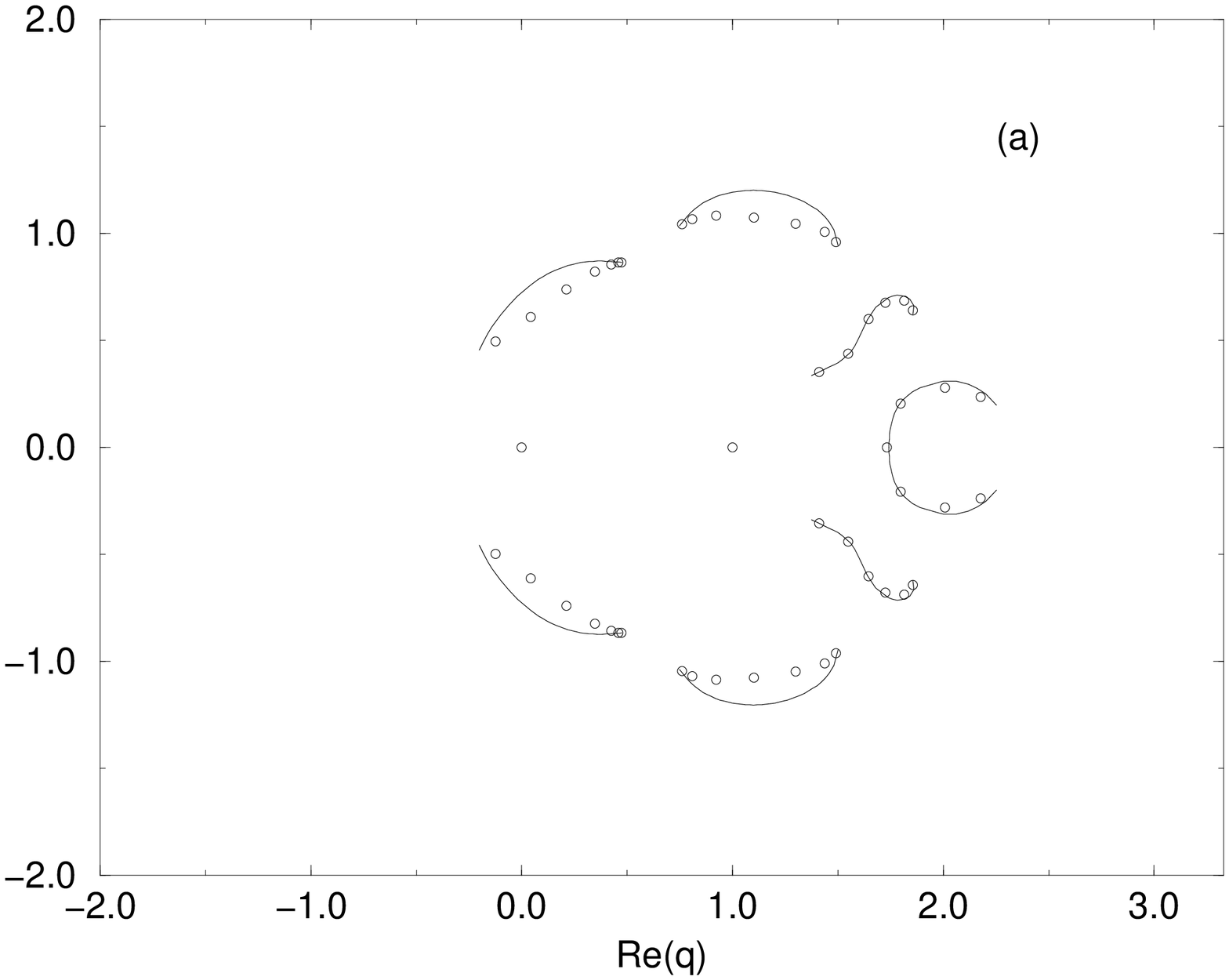}
\end{center}
\vspace{-3cm}
\begin{center}
\leavevmode
\epsfxsize=3.0in
\epsffile{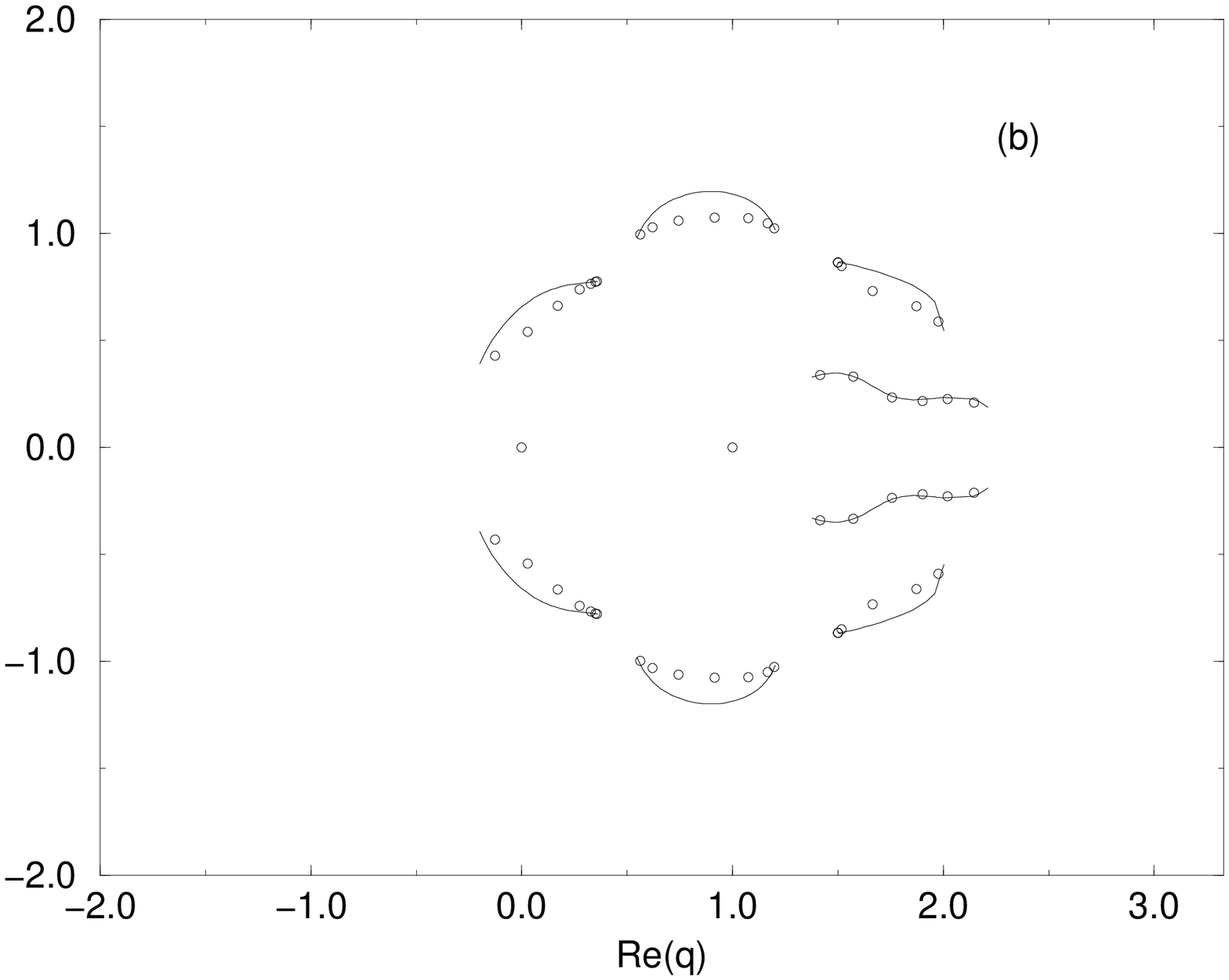}
\end{center}
\vspace{-2cm}
\caption{\footnotesize{As in Fig. \ref{hep3k34} for $p=4$ and $k=$ (a) 7 (b)
8.  For comparison, chromatic zeros are shown for $m=6$.}}
\label{hep4k78}
\end{figure}

\pagebreak

\section{General Properties of $W$ and ${\cal B}$}

We note several general features. In the following it is understood that 
(i) $k \ge 3$, since ${\cal B} = \emptyset$ for $k=2$; and 
(ii) $p \ge 3$, since we have already analyzed the $p=2$ case in detail 
\cite{wa23}. 

\begin{enumerate}

\item 

For all $q \in {\cal B}$, $|q|$ is bounded above, i.e., the loci ${\cal B}$ 
are compact and do not extend infinitely far from the origin, $q=0$. 

\item 

For a given $(p,k)$, the nonanalytic locus ${\cal B}$ in the $L_m$ limit of
eq. (\ref{minf}) is comprised of arcs and, if and only if $p$ is odd, also a 
line segment on the positive real $q$ axis.  These arcs and possible line
segment do not enclose any regions in the $q$ plane.

\item

For a number of values of $p$ and $k$, ${\cal B}$ includes support for 
$Re(q) < 0$, although it never includes a line segment on, or a complex arc
crossing or touching, the negative real axis. 
Typically for a given $p$, as one increases $k$ beyond $k=3$ to moderate
values, the left-hand endpoints of the left-most complex-conjugate arcs move 
into the $Re(q) < 0$ half-plane.  Since these complex-conjugate arcs with
left-hand endpoints extending into the $Re(q) < 0$ half-plane intersect the 
imaginary axis in the $q$ plane away from the origin, the loci 
${\cal B}$ also have the property that $Re(q) = 0$ does not necessarily 
imply that $q=0$.  These statements about ${\cal B}$ also apply to chromatic
zeros for finite-length strips, as is evident from the figures. 

\item  

For even $p$, the number of different components (arcs) on ${\cal B}$ is
\beq
N_{comp.} = k+p-4  \quad {\rm for \ even} \quad p
\label{ncpeven}
\eeq
Since no two of these arcs touch each other, the number of arc \underline 
endpoints is 
\beq
N_e = 2N_{comp.} = 2(k+p-4) \quad {\rm for \ even} \quad p
\label{nepeven}
\eeq
If $p$ is even and (i) $k$ is even, then these arcs form $(k+p-4)/2$
complex-conjugate pairs, none of which crosses the real axis, so 
\beq
p \ even \ , \ \ k \ even \quad \Longrightarrow \quad {\rm no} \quad q_c
\label{pevenkevennoqc}
\eeq
(ii) $k$ is odd, then there are $(k+p-5)/2$
complex-conjugate pairs of arcs and, in addition, a self-conjugate arc crossing
the positive real axis at the point $q_c$.  ${\cal B}$ does not contain any
line segment on the real $q$ axis.  The point $q_c$ monotonically increases as 
$k$ increases through odd values for fixed even $p$ and, separately, as $p$
increases through even values for fixed odd $k$.  We have 
\beq
\lim_{k \ odd \ \to \infty}q_c(p \ even) = 
\lim_{p \ even \to \infty}q_c(k \ odd)= 2
\label{pevenqckinf}
\eeq
For example, we have 
\beq
q_c(p=4,k=3)=1.56984...
\label{qcp4k3}
\eeq
\beq
q_c(p=4,k=5)=1.68233...
\label{qcp4k5}
\eeq
\beq
q_c(p=6,k=3)=1.59750...
\label{qcp6k3}
\eeq
\beq
q_c(p=6,k=5)=1.70074...
\label{qcp6k5}
\eeq
Since $\chi=2$ for even $p$, from eq. (\ref{chi}), it follows that for finite 
odd $k$, where there is a $q_c$ defined, $q_c < \chi$. 

\item 

For odd $p$, the number of different components on ${\cal B}$ is
\beq
N_{comp.} = 2 \Bigl [ \frac{k}{2} \Bigr ]_i+p-4  \quad {\rm for \ odd} \quad p
\label{ncpodd}
\eeq
where $[x]_i$ denotes the integral part of $x$.  The locus ${\cal B}$ 
always includes a line segment on the positive real $q$ axis given
by 
\beq
q_1 \le q \le q_2=2
\label{linesegment}
\eeq
Since $q_2$ is the maximal finite real point where $W$ is nonanalytic, we have 
\beq
q_c(p \ odd; \ any \ k) =2
\label{qc2}
\eeq
Since by eq. (\ref{chi}), the chromatic number is $\chi=3$ for odd $p$, it
follows that 
\beq
q_c = \chi -1 \quad {\rm for \ odd} \quad p 
\label{qcchim1}
\eeq
For a given odd value of $k$, the left-hand endpoint of the line segment, 
$q_1$, is a monotonically increasing function of $p$ and approaches the limit 
\beq
\lim_{p \ odd \ \to \infty} q_1=2 \quad {\rm for \ any} \quad k
\label{q1pinf}
\eeq 
so that in this limit the line segment shrinks to zero and disappears.
For a given odd value of $p$, as $k$ increases through odd (even) values from 
3 \ (4) toward infinity, the value of the lower endpoint increases (decreases) 
monotonically toward a limiting value, $q_1(k=\infty)$.  
Some examples of these limiting values
are 
\beq
q_1(p=3,k=\infty) = \frac{4}{3}
\label{q1kinfp3}
\eeq
\beq
q_1(p=5,k=\infty) = \frac{3}{2}
\label{q1kinfp5}
\eeq
\beq
q_1(p=7,k=\infty) = 1.589755..
\label{q1kinfp7}
\eeq
\beq
q_1(p=9,k=\infty) = 1.647799..
\label{q1kinfp9}
\eeq
In general, $q_1(p \ fixed,k=\infty)$ is a monotonically increasing function 
of $p$ and
approaches the limit (\ref{q1pinf}) as $p \to \infty$.  Some specific values of
$q_1$ for odd $p$ and finite $k$ include 
\beq
q_1(p=3,k=3)=1.29560..
\label{q1p3k3}
\eeq
\beq
q_1(p=3,k=4)=1.35321..
\label{q1p3k4}
\eeq
\beq
q_1(p=5,k=3)=1.43016..
\label{q1p5k3}
\eeq
\beq
q_1(p=5,k=4)=1.56984..
\label{q1p5k4}
\eeq
If $k$ is even, no two of the components of ${\cal B}$ touches any other, so
that the number of endpoints of the arcs and the line segment is given by 
\beq
N_e = 2N_{comp.} = 2(k+p-4) \quad {\rm for \ odd} \quad p \quad 
{\rm and \ even} \quad k
\label{neoddpoddk}
\eeq
If $k$ is odd, then two of the complex-conjugate arcs have right-hand ends that
meet the right-hand side of the line segment at $q_c=2$ to form a multiple
point (in the technical terminology of algebraic geometry \cite{mp}) 
with valency 3 on 
the algebraic curve constituted by ${\cal B}$.  Hence, with $N_{comp.}=k+p-5$
from eq. (\ref{ncpodd}), we have, for the number of endpoints on ${\cal B}$, 
\beq
N_e = 2(N_{comp.}-1)+3=2k+2p-9 \quad {\rm for \ odd} \quad p \quad 
{\rm and \ odd} \quad k
\label{n3oddpevenk}
\eeq
In both of these cases of even and odd $k$, the endpoints and the odd--$k$
multiple point are located at zeros of $R$ and hence the branch points of 
$\sqrt{R}$ in eq. (\ref{lambda12}). 

\end{enumerate}

\vspace{8mm}

One can understand these observed properties of ${\cal B}$ as follows.  As
before, we restrict our consideration to the cases that we have not already
studied, namely $p \ge 3$ and $k \ge 3$.  The boundedness of $|q|$ for 
$q \in {\cal B}$ follows because the degeneracy equation eq. (\ref{degeneq})
does not satisfy the necessary and sufficient condition that we gave in
(section IV of) Ref. \cite{wa} for ${\cal B}$ to extend infinitely far from 
the origin, $q=0$.  Now for large $|q|$ (and $p \ge 3$), the
degree in $q$ of the two highest--power terms in $\lambda_{1,2}$ are the same:
\beq
deg_q((q-1)D_{k+p-3}) = deg_q(\sqrt{R}) = k+p-4
\label{degreeeq}
\eeq
where the term in $R$ with the highest-power in $q$ is
simply $[(q-1)D_{k+p-3}]^2$.  Hence, for $\lambda_1$, where
$\sqrt{R}$ enters with a $+$ sign, these highest-power terms add, so 
$deg_q(\lambda_1)=k-p-4$, 
while for $\lambda_2$, where $\sqrt{R}$ enters with a 
$-$ sign, they cancel, so the resultant leading power of $q$ for large $|q|$ is
smaller.  Since the leading powers in $q$ are different for $\lambda_1$ and
$\lambda_2$, it follows that the degeneracy equation cannot be satisfied for
arbitrarily large $|q|$ and hence ${\cal B}$ is bounded in the $q$ plane. 
The reasoning underlying the structure of ${\cal B}$ as a set of arcs and 
the line segment is the same as we gave in our analysis of strip graphs in
Ref. \cite{strip}.  This also explains the fact that the endpoints of the
arcs are the branch points of the term $\sqrt{R}$ occurring in $\lambda_{1,2}$.

The third feature that we have observed is of interest since for many simple
families of graphs that we have studied \cite{w,wc,wa23}, including all of the
infinitely long, finite-width strip graphs analyzed in Ref. \cite{strip}, the 
respective nonanalytic loci ${\cal B}$ have the property that if $q \in 
{\cal B}$, then $Re(q) \ge 0$ and the further property that if 
$q \in {\cal B}$ and $Re(q)=0$, then $q=0$. We had previously encountered 
loci ${\cal B}$ that included support for $Re(q) < 0$ and for which $Re(q)=0$ 
did not imply that $q=0$ when studying families with noncompact ${\cal B}$'s 
that extended infinitely far from the origin $q=0$ \cite{wa23}.  
However, here we observe this phenomenon for simple families
with compact loci ${\cal B}$ that are bounded in the $q$ plane. This behavior
of the asymptotic loci is also reflected in the chromatic zeros for finite-$n$
graphs, as is evident from the figures.  Let us denote the minimum value of 
$Re(q)$ for $q \in {\cal B}$ as 
\beq
(q_{_R})_{min} \equiv min[Re(q \in {\cal B})]
\label{qrmin}
\eeq
As is evident from the figures, 
for $p=3$, $(q_{_R})_{min}$ has the values 0.3522 for $k=3$, 
$-0.1766$ for $k=4$, with similar negative values for other $k$ that
we have studied, while for $p=4$, $(q_{_R})_{min}$ has the values 
0.2047 for $k=3$, $-0.0813$ for $k=4$, with similar negative values for 
other $k$
that we have studied.  As will be discussed further below, as 
$k \to \infty$ for fixed $p \ge 3$ or as $p \to \infty$ for fixed 
$k \ge 3$, some arcs tend to approach the unit circle $|q-1|=1$, so the motion
of the left endpoints of the leftmost arcs in ${\cal B}$ is not expected to be
a monotonic function of $k$ for fixed $p$ or of $p$ for fixed $k$. 

Note that 
both the families in Ref. \cite{wa23} with noncompact ${\cal B}$'s and the
families with compact ${\cal B}$'s including support for $Re(q) < 0$ in the
present work are consistent with our earlier conjectures (1a) and (1b) in Ref. 
\cite{strip}.  We recall that conjecture (1a) stated that if $q_0$ is a
chromatic zero of a regular lattice graph $G$ with no global circuits 
\cite{global}, then $Re(q_0) \ge 0$ and, furthermore, the only chromatic zero
with $Re(q_0)=0$ is $q_0=0$ itself.  Conjecture (1b) stated that if one
considers the $n =v(G) \to \infty$ limit of a regular lattice graph $G$ with no
global circuits, then the points $q \in {\cal B}$ have the property that 
$Re(q) \ge 0$, and furthermore, the only point on ${\cal B}$ with $Re(q) = 0$
is the point $q=0$ itself.  It is true that our homeomorphic expansions of
strip graphs do not contain any global circuits.  However, our results are
still consistent with the conjectures in Ref. \cite{strip} because, as 
discussed in Ref. \cite{wa23}, none of the families analyzed there was a
regular lattice graph, and this is similarly true of the homeomorphic families
that exhibit chromatic zeros and ${\cal B}$ with $Re(q) < 0$ in the present
work, i.e., those with $p \ge 3$ and $k \ge 3$.  This is clear since the 
definition of a regular
lattice graph requires all vertices except possibly those on the boundary to
have the same degree (= coordination number).  But all of the families with 
$p \ge 3$ and $k \ge 3$ involve homeomorphic expansion and hence have 
vertices with degree 2 as well as vertices with the usual degree for the 
given lattice (and boundary vertices).  Hence, these homeomorphic expansions 
are not regular lattice graphs. 

For the other properties in the list above, it is necessary to deal with the
even--$p$ and odd--$p$ cases separately.  The degree of the discriminant $R$,
as a polynomial in $q$, is $2(k+p-4)$ from eq. (\ref{degqr}), and (aside
from the cases $p=2$ and $k=2$) $R$ has only simple zeros, so that each of
these corresponds to a branch point of $\sqrt{R}$.  If $p$ and $k$ are
both even, $R$ has no real zeros; i.e., all branch points of $\sqrt{R}$ occur
in complex-conjugate pairs, connected by the arcs of ${\cal B}$.  This yields
the results in eqs. (\ref{ncpeven}) and (\ref{nepeven}).  If $p$ is
even but $k$ is odd, then $R$ has no real zeros while $b_1$ has a single
real zero, so that at this zero, $\lambda_1=(1/2)\sqrt{R}=-\lambda_2$, which
evidently satisfies the degeneracy equation (\ref{degeneq}).  This explains why
for this case, $p$ even and $k$ odd, ${\cal B}$ contains a self-conjugate arc
that crosses the real axis.  The zero of $b_1$ is precisely $q_c$, and this
has the monotonicity properties discussed above, approaching the limit 
$q_c=2$ as $p \to \infty$ through even values for fixed odd $k$ or 
$k \to \infty$ through odd values for fixed even $p$, as in eq.
(\ref{pevenqckinf}).  The other arcs on ${\cal B}$ form 
complex-conjugate pairs away from the real axis.  

  For odd $p$, we can explain the existence of the line segment 
that occurs on the real $q$ axis by noting that in this case,
for $q \in {\mathbb R}$, since the first two terms in $\lambda_{1,2}$ are
real and nonzero, 
there will be such a portion of ${\cal B}$ on the real axis if and only
if $R < 0$ so that $\sqrt{R}$ is pure imaginary, so that $\lambda_1 =
r+ir'$, which is equal in magnitude to $\lambda_2 = r-ir'$, where $r$ and 
$r'$ denote real quantities.  Now eq. (\ref{degqr}) shows that the degree in
$q$ of $R$ is even, so that $R > 0$ for large positive and negative $q$.  For
odd $p$, $R$ has two simple zeros at the positive real points $q_1$ and 
$q_2$, so that $R < 0$ for $q_1 < q < q_2$, which is the requisite condition 
for the line segment \cite{pol}.  Furthermore, $q_2=2$ for arbitrary $k$. This
is a consequence of the fact that $b_1$ contains a factor of $(q-2)$ by eq. 
(\ref{dk2shift}) and $b_2 \propto D_kD_{k-1}$ also contains a factor of 
$(q-2)$ by eq. (\ref{dkkoddfactor}) since either $k$ or $k-1$ is odd.  The 
values of $q_1$ follow from a straighforward analysis of the $R=0$ equation;
for example, the values of $q_1$ for the limit $k \to \infty$ are obtained 
from the equation $(1-a)^2-4a^{p-1}=0$ where $a=q-1$; since $p=2\ell+1$ is 
odd, this can be factorized as $[1-a-2a^\ell][1-a+2a^\ell]=0$.  
Because $p$ is odd, it follows from the
general eqs. (\ref{chi}) and (\ref{pq2zeropodd}) that 
$\chi((Ch)_{p \ odd,k,m})=3$ and hence $P((Ch)_{p \ odd,k,m},q=2)=0$ as in 
eq. (\ref{pq2zeropodd}).  {\it A priori }, 
a zero of a chromatic polynomial $P(G,q)$ at some value $q=q_0$
is not necessarily also a zero of the asymptotic function $W$; this is true 
if and only if the multiplicity of this zero in $P$ is proportional to the
number of vertices $n=v(G)$ as $n \to \infty$.  This is, indeed, true of 
the zero at $q=2$ in eq. (\ref{pq2zeropodd}), as follows from our previous
demonstration that for odd $p$, both $b_1$ and $b_2$, and hence also 
$\lambda_{1,2}$, contain a factor of $q-2$.  Hence, 
\beq
W([\lim_{m \to \infty}(Ch)_{p \ odd,k,m}],q) \propto 
(q-2)^{\frac{1}{2(k+p-4)}}
\label{wqm2factor}
\eeq

We can obtain an additional result for the case $p=3$.  In this case, from 
eqs. (\ref{lambda12}) and (\ref{disc}), it follows
that each term proportional to $x$ in ${\cal N}((Ch)_{3,k},q,x)$ and 
${\cal D}((Ch)_{3,k},q,x)$ contains a factor of $D_k$, and 
$\lambda_{1,2}$ contains an overall factor of $(D_k)^{1/2}$, since 
\beq
\lambda_{1,2}(p=3) = \frac{1}{2}\biggl [ (q-2)D_k \pm \sqrt{R} \biggr ]
\label{lam12ch12kp3}
\eeq
where 
\beq
R(p=3) = D_k\biggl [ (q-2)^2D_k + 4(q-1)^2D_{k-1} \biggr ]
\label{rch12kp3}
\eeq
Hence, for the nontrivial case $m \ge 1$, 
\beq
P((Ch)_{3,k,m},q) = 0 \quad {\rm at \ the \ zeros \ of} \quad D_k
\label{pch12kmdkzero}
\eeq
and, indeed, each of these zeros of the chromatic polynomial have multiplicity
proportional to $n$, so that 
\beq
W([\lim_{m \to \infty}(Ch)_{3,k,m}],q) \propto (D_k)^{\frac{1}{2(k-1)}}  
\label{wdkfactor}
\eeq
and 

\vspace{8mm}

As $k$ gets large for fixed $p$, one can see from the figures that some arcs 
are located close to the unit circle $|q-1|=1$, while for $p \ge 4$, some 
arcs are oriented almost orthogonally to this circle, and, for odd $p$, the 
line segment on the real axis approaches the limit
\beq
q_1(k=\infty) \le q \le 2
\label{linesegmentkinf}
\eeq
where specific values of the lower endpoint, $q_1(k=\infty)$, were given above
for various values of odd $p$.  

One can also consider the limit as $p$ gets large for fixed $k$; here one sees
an alternation of features such as the presence (absence) of the line segment 
for odd (even) $p$.  However, as noted above (cf. eq. (\ref{q1pinf})), since 
$\lim_{p \ odd \ \to \infty}q_1=2$, the line segment disappears in this limit.
Comparing the figures with different $p$ but the same $k$, one notices that
some of the arcs tend to approach the unit circle $|q-1|=1$.

Another property that we observe concerns the zeros of the chromatic
polynomials (=chromatic zeros) for finite--$m$ graphs $(Ch)_{p,k,m}$. For most
of the complex arcs the chromatic zeros lie close to, but not, in general, 
exactly on, these asymptotic curves.  In contrast, for odd--$p$ families, 
where ${\cal B}$ contains a line segment on the positive real axis, 
$q_1 \le q \le 2$, the
chromatic zeros lie exactly on this segment, i.e., are real. {\it A priori},
one could have envisioned a different situation in which some chromatic zeros
that have real parts $Re(q)$ roughly in the range $q_1 \le Re(q) \le 2$ also
have imaginary parts that approach zero as $m \to \infty$ so that they move 
in to the real axis vertically or obliquely from above and below to form 
(part or all of) the line segment.  But instead, the
chromatic zeros that merge to form the line segment are real even for finite
strips.  One curious feature that we observe is that in certain cases even 
for complex arcs, the chromatic zeros lie very close to, or even (to within 
machine precision) on the asymptotic curves, such as for the northeast and
southeast arcs in Fig. \ref{hep5k34}(b) for $p=5$, $k=4$ and the middle,
vertically oriented arcs in Fig. \ref{hep5k56}(a) for $p=5$, $k=5$, and 
Fig. \ref{hep6k56}(a,b) for $p=6$, $k=5,6$, as well as the complex 
self-conjugate arcs in Fig. \ref{hep6k34}(a) for $p=6$, $k=3$
and Fig. \ref{hep6k56}(a) for $p=6$, $k=5$.  Earlier we had encountered similar
behavior for certain families of graphs \cite{w} and proved a general theorem
that showed this rigorously for a class of graphs that we called ``$p$-wheels''
\cite{wc}.  As future work, it would be worthwhile to try to construct an 
analogous proof in the present case.

Finally, we observe that ${\cal B}$ for $\lim_{m \to \infty}(Ch)_{p=4,k=3,m}$,
shown in Fig. \ref{hep4k34}(a)
is related to the locus ${\cal B}$ that we calculated previously for the
infinitely long open strip of the triangular lattice of width $L_y=3$
\cite{strip} (see Fig. Fig. 5(a) of Ref. \cite{strip}) 
by a simple horizontal translation by 1 unit in the $q$ plane: 
\beq
{\cal B}(q)_{\lim_{m \to \infty}(Ch)_{p=4,k=3,m}} = 
{\cal B}(q')_{G_{t(L_y=3)}} \quad {\rm for} \quad q'=q+1
\label{brel}
\eeq
This is a consequence of the fact that 
\beq
{\cal D}((Ch)_{p=4,k=3},q,x) = {\cal D}(G_{t(L_y=3)},q',x) \quad {\rm for}
\quad q'=q+1
\label{drel}
\eeq
so that 
\beq
\lambda_{(Ch)_{p=4,k=3},j}(q)=\lambda_{G_{t(L_y=3)},j}(q') , \ j=1,2 
\quad {\rm for} \quad q'=q+1
\label{lambdarel}
\eeq
(This relation does not hold for the respective numerator functions $A_0$ and
$A_1$, but these do not affect ${\cal B}$.) 

\section{Comparison of Loci ${\cal B}$ in the $L_{\lowercase{m}}$, 
$L_{\lowercase{p}}$, and $L_{\lowercase{k}}$ Limits}

As discussed in the introduction, there are three different ways of 
sending the number of vertices of the graph $(Ch)_{p,k,m}$ to infinity, as
specified in the three limits $L_m$, $L_p$, and $L_k$, eqs. 
(\ref{minf})-(\ref{kinf}).  We have concentrated on the $L_m$ limit because it
is the most interesting and is a generalization of our previous study of
infinitely long, finite-width strip graphs of various lattices in Ref. 
\cite{strip}.  In this section we present the analogous, and much simpler
results for ${\cal B}$ and $W$ in the other two limits.  We recall first that
for both the $L_p$ and $L_k$ limit, the trivial case $m=0$ leads to ${\cal B} =
\emptyset$ so we take $m \ge 1$.  Furthermore, in the $L_p$ limit, as was true
for the $L_m$ limit, the case $k=2$ is trivial (cf. eq. (\ref{pchainmpgons}))
and again yields ${\cal B}=\emptyset$, so for the $L_p$ limit we take 
$k \ge 3$.  In passing we note that whereas for the 
$L_m$ limit, the value $p=2$ leads to a noncompact ${\cal B}(q)$, extending
infinitely far from the origin, $q=0$, which we have analyzed in detail
previously \cite{wa23}, for the $L_k$ limit, ${\cal B}$ is the same for $p=2$
and $p \ge 2$ so we include all $p \ge 2$ in this case. 

 From our general results for $P((Ch)_{p,k,m},q)$, we find that 
in the $L_k$ limit, $k \to \infty$ for fixed $m \ge 1$ and $p \ge 2$, and,
separately, in the 
$L_p$ limit, $p \to \infty$ for fixed $m \ge 1$ and $k \ge 3$, 
\beq
{\cal B}_{L_k} = {\cal B}_{L_p} = q: \quad |q-1|=1 
\label{blklp}
\eeq
Clearly, there 
is a fundamental topological difference between ${\cal B}$ for the $L_k$ and 
$L_p$ limits on the one hand, and the respective ${\cal B}$'s for the $L_m$ 
limit with various values of $p \ge 3$ and $k \ge 2$ on the other hand: 
the ${\cal B}$'s for the $L_k$ and $L_p$ limits divide the complex
$q$ plane into two regions, whereas the ${\cal B}$'s for the $L_k$ limit with
various $p$ and $k$ do not.  Another important
difference is that ${\cal B}$ in the $L_k$ limit is actually independent of the
specific values of the other parameters, $m$ and $p$, provided that they
satisfy $m \ge 1$ and $p \ge 2$, and, similarly, ${\cal B}$ in the $L_p$ 
limit is independent of the specific values of $m$ and $k$, provided that they
satisfy $m \ge 1$ and $k \ge 3$; in contrast, 
the locus ${\cal B}$ in the $L_m$ limit does depend on the specific values of 
$p$ and $k$ provided that $p \ge 2$ and $k \ge 3$.

  Defining, as in our earlier work \cite{w,wc,wa,wa23}, the 
region $R_1$ as the region including the positive real
axis for large $q$ and extending down to the maximal finite real point, $q_c$,
where $W$ is nonanalytic, we identify
\beq
q_c = 2 \quad {\rm for} \quad {\cal B}_{L_k}, \ {\cal B}_{L_p}
\label{qcblklp}
\eeq
and
\beq
R_1 = q: \ |q-1| > 1 \quad {\rm for} \quad {\cal B}_{L_k}, \ {\cal B}_{L_p}
\label{r1lklp}
\eeq
The other region is the complement of $R_1$, 
\beq
R_2 = q: \ |q-1| < 1 \quad {\rm for} \quad {\cal B}_{L_k}, \ {\cal B}_{L_p}
\label{r2lklp}
\eeq
For the nontrivial case $m \ge 1$ and the respective limits 
$L_k$ with $p \ge 2$ and $L_p$ with $k \ge 3$, we find 
\beq
W((Ch)_{p,k,m};L_k,q) = W((Ch)_{p,k,m};L_p,q) = q-1 \quad {\rm for} \quad 
q \in R_1
\label{wr1blklp}
\eeq
As we have discussed before \cite{w}, in regions other than $R_1$, one can only
determine the magnitude of $W$; we obtain
\beq
|W((Ch)_{p,k,m};L_k,q)| = |W((Ch)_{p,k,m};L_p,q)| = 1 \quad {\rm for} \quad 
q \in R_2
\label{wr2blklp}
\eeq

\begin{figure}
\vspace{-4cm}
\centering
\leavevmode
\epsfxsize=3.0in
\begin{center}
\leavevmode
\epsffile{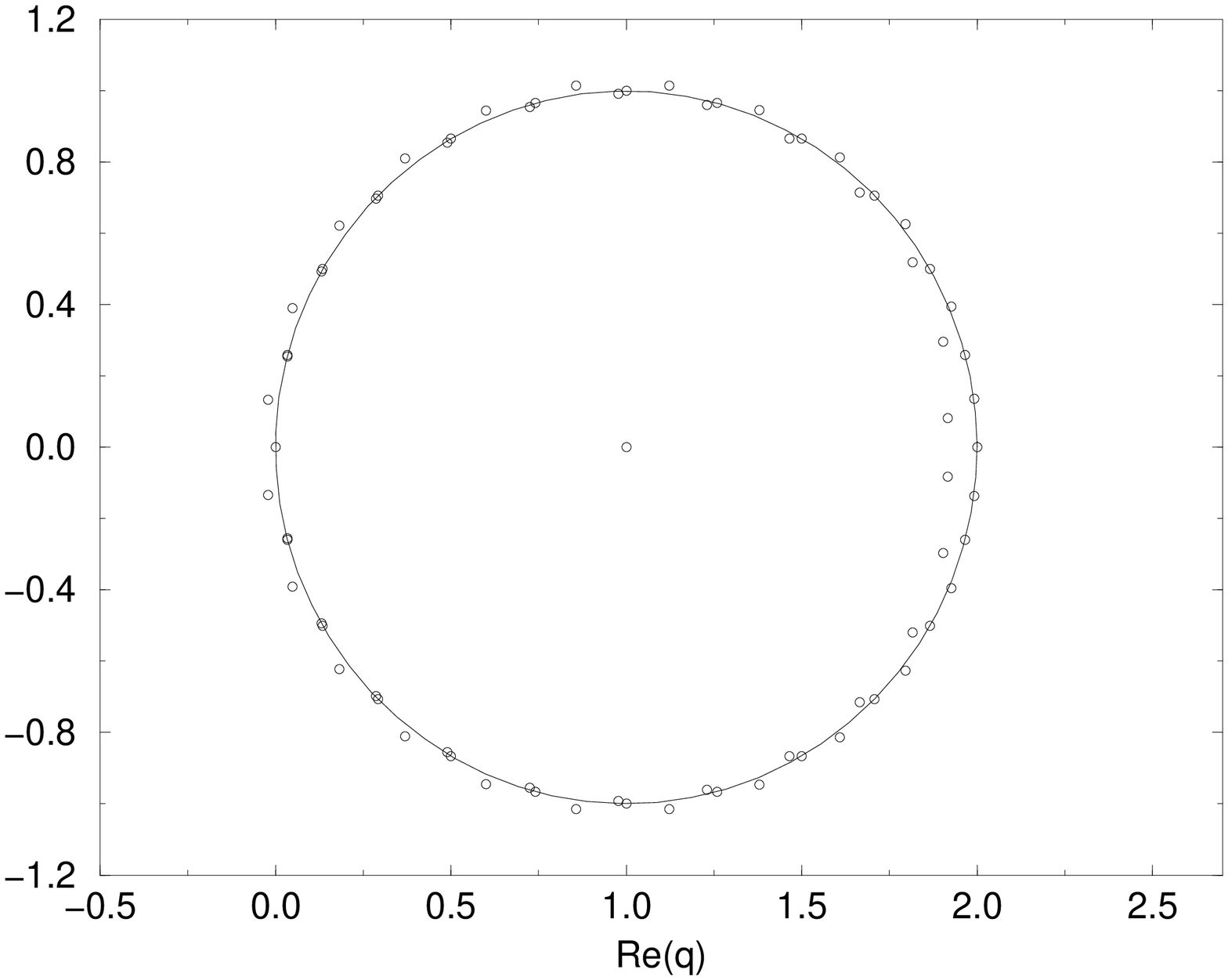}
\end{center}
\vspace{-2cm}
\caption{\footnotesize{Locus ${\cal B}$ and analytic structure of the
function $W([\lim_{k \to \infty}(Ch)_{p,k,m}],q)$ for $p=3$ and $m=2$. 
For comparison, chromatic zeros are shown for $k=25$.}}
\label{hep3k25m2q}
\end{figure}

\begin{figure}
\vspace{-4cm}
\centering
\leavevmode
\epsfxsize=3.0in
\begin{center}
\leavevmode
\epsffile{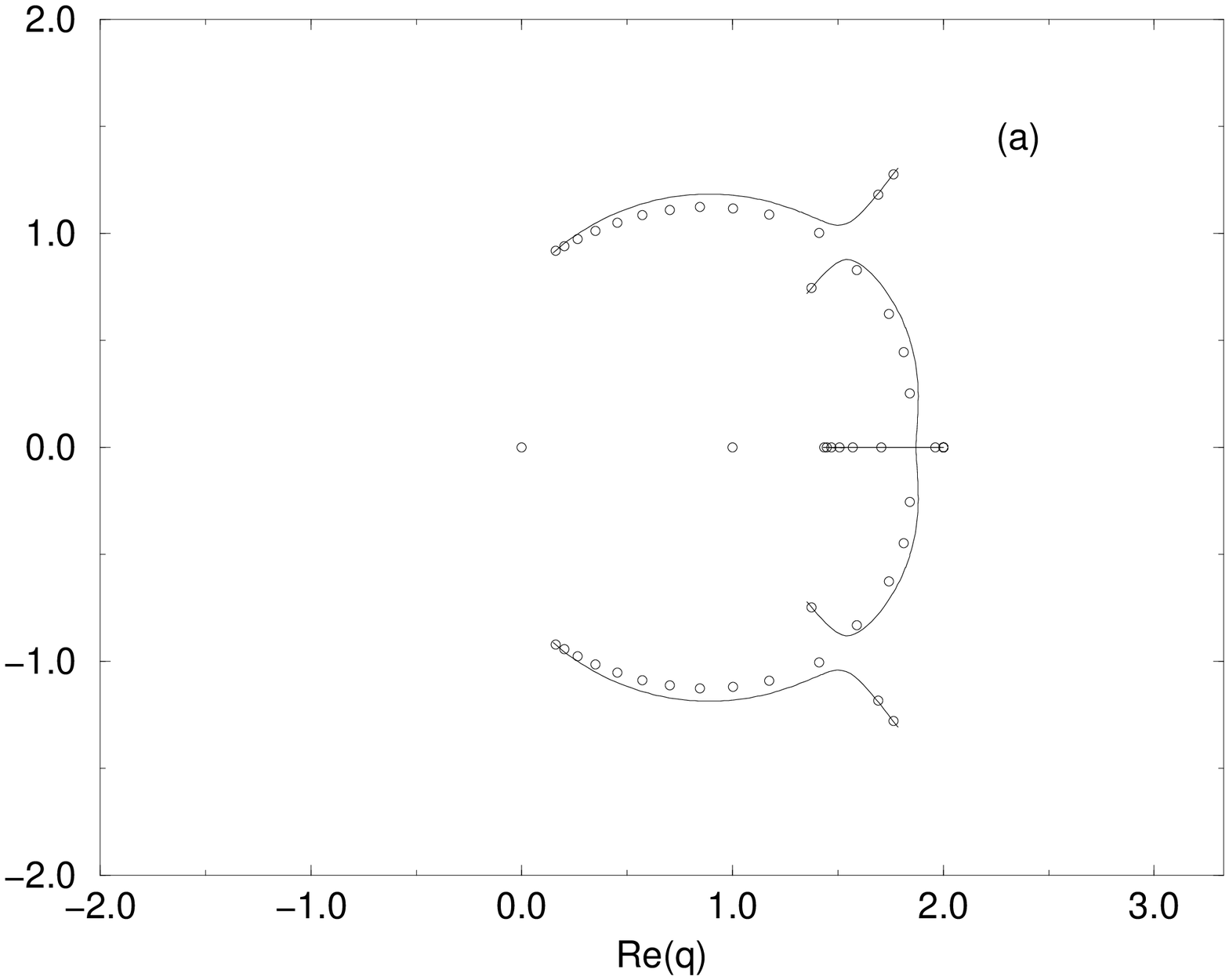}
\end{center}
\vspace{-3cm}
\begin{center}
\leavevmode
\epsfxsize=3.0in
\epsffile{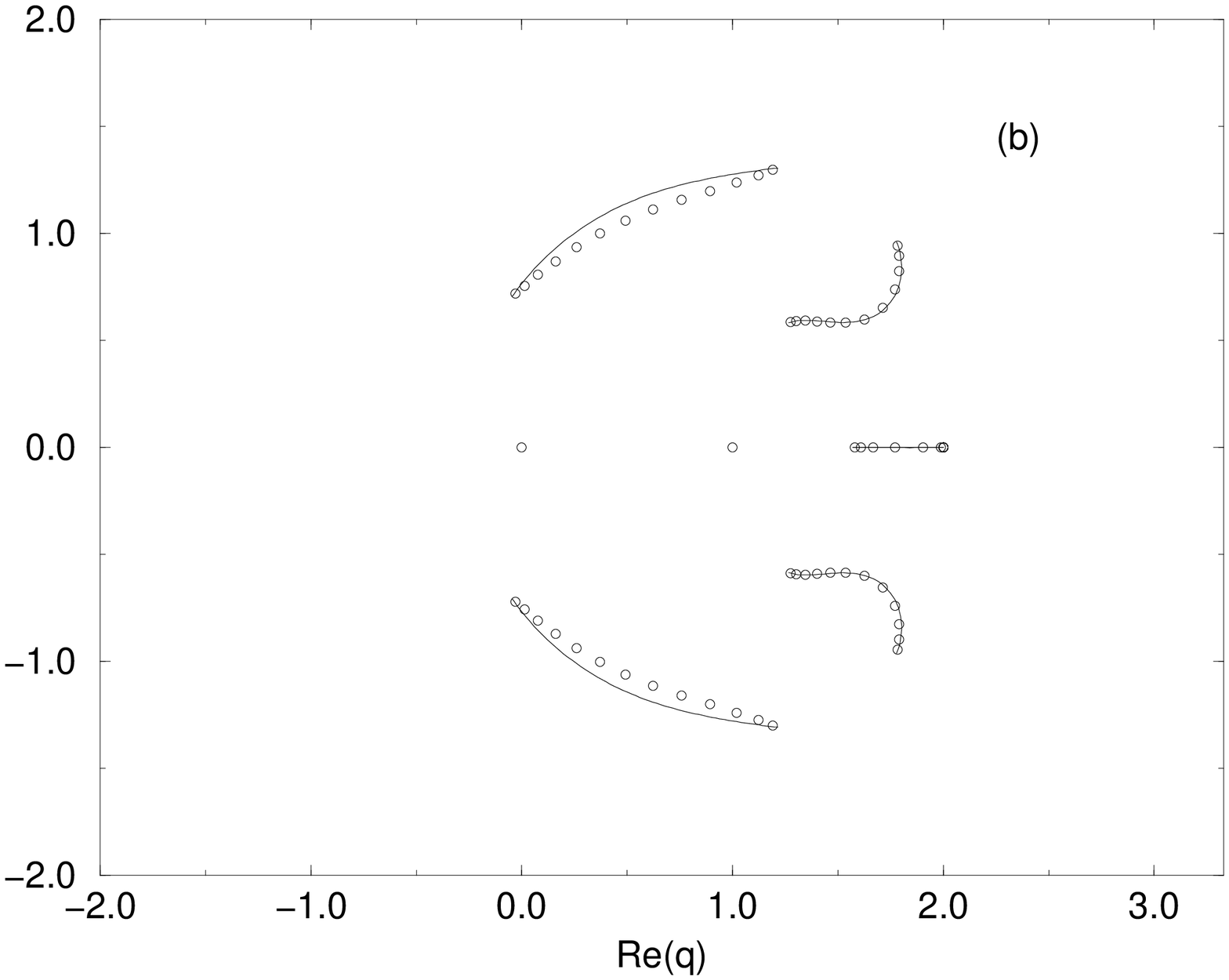}
\end{center}
\vspace{-2cm}
\caption{\footnotesize{As in Fig. \ref{hep3k34} for $p=5$ and $k=$ (a) 3 (b)
4.  For comparison, chromatic zeros are shown for $m=12$.}}
\label{hep5k34}
\end{figure}

\begin{figure}
\vspace{-4cm}
\centering
\leavevmode
\epsfxsize=3.0in
\begin{center}
\leavevmode
\epsffile{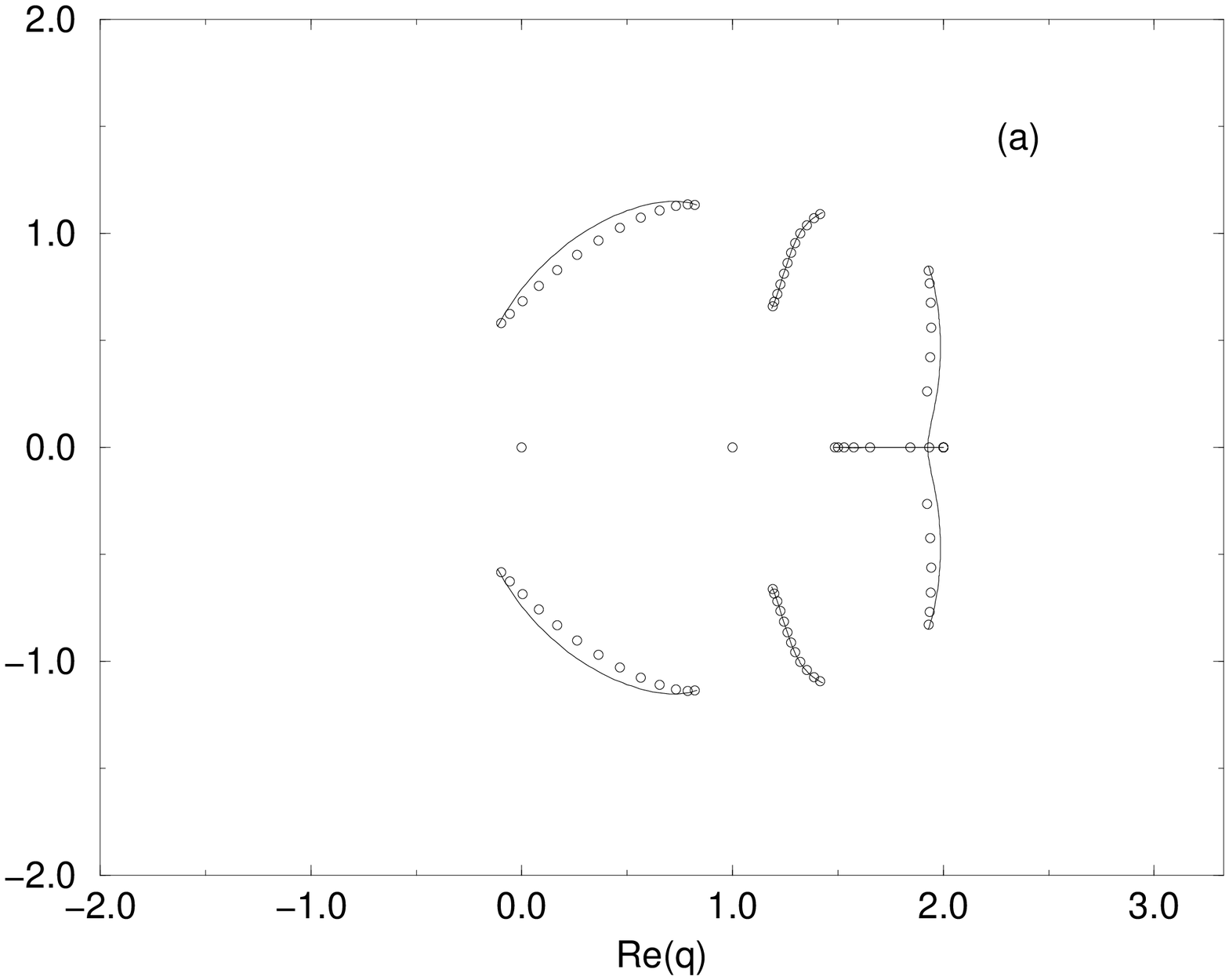}
\end{center}
\vspace{-3cm}
\begin{center}
\leavevmode
\epsfxsize=3.0in
\epsffile{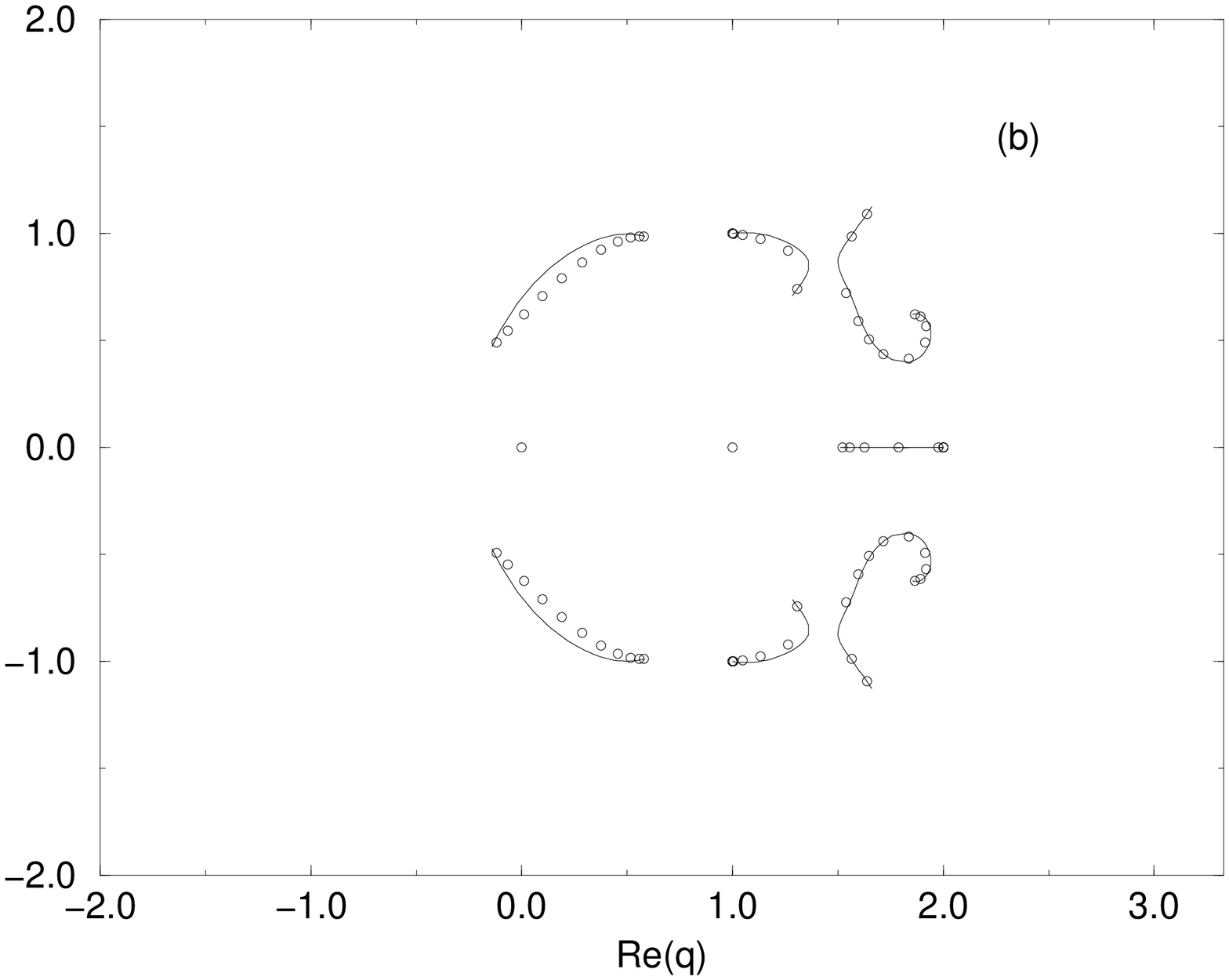}
\end{center}
\vspace{-2cm}
\caption{\footnotesize{As in Fig. \ref{hep3k34} for $p=5$ and $k=$ (a) 5 (b)
6.  For comparison, chromatic zeros are shown for $m=$ (a) 12 (b) 10.}}
\label{hep5k56}
\end{figure}

\begin{figure}
\vspace{-4cm}
\centering
\leavevmode
\epsfxsize=3.0in
\begin{center}
\leavevmode
\epsffile{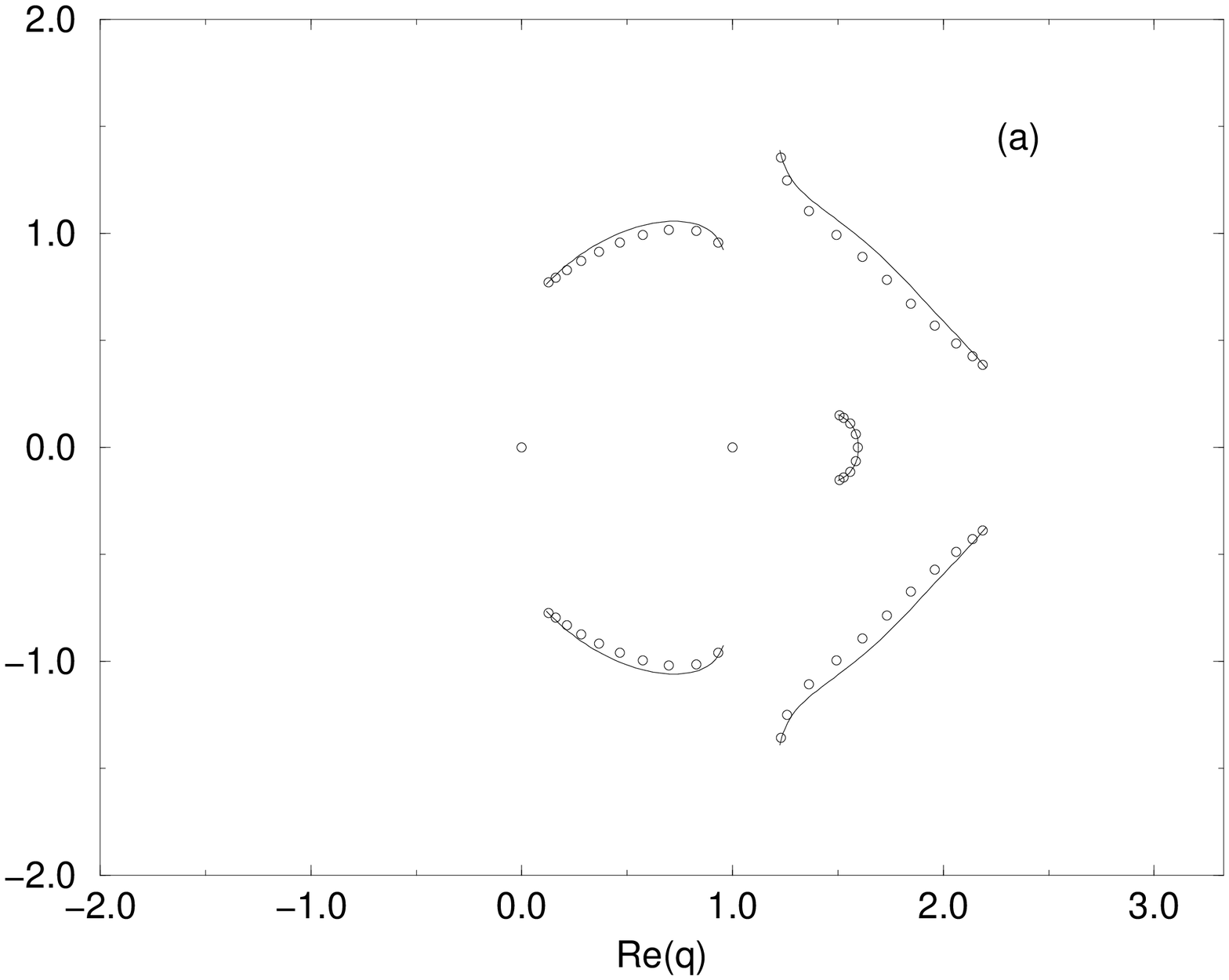}
\end{center}
\vspace{-3cm}
\begin{center}
\leavevmode
\epsfxsize=3.0in
\epsffile{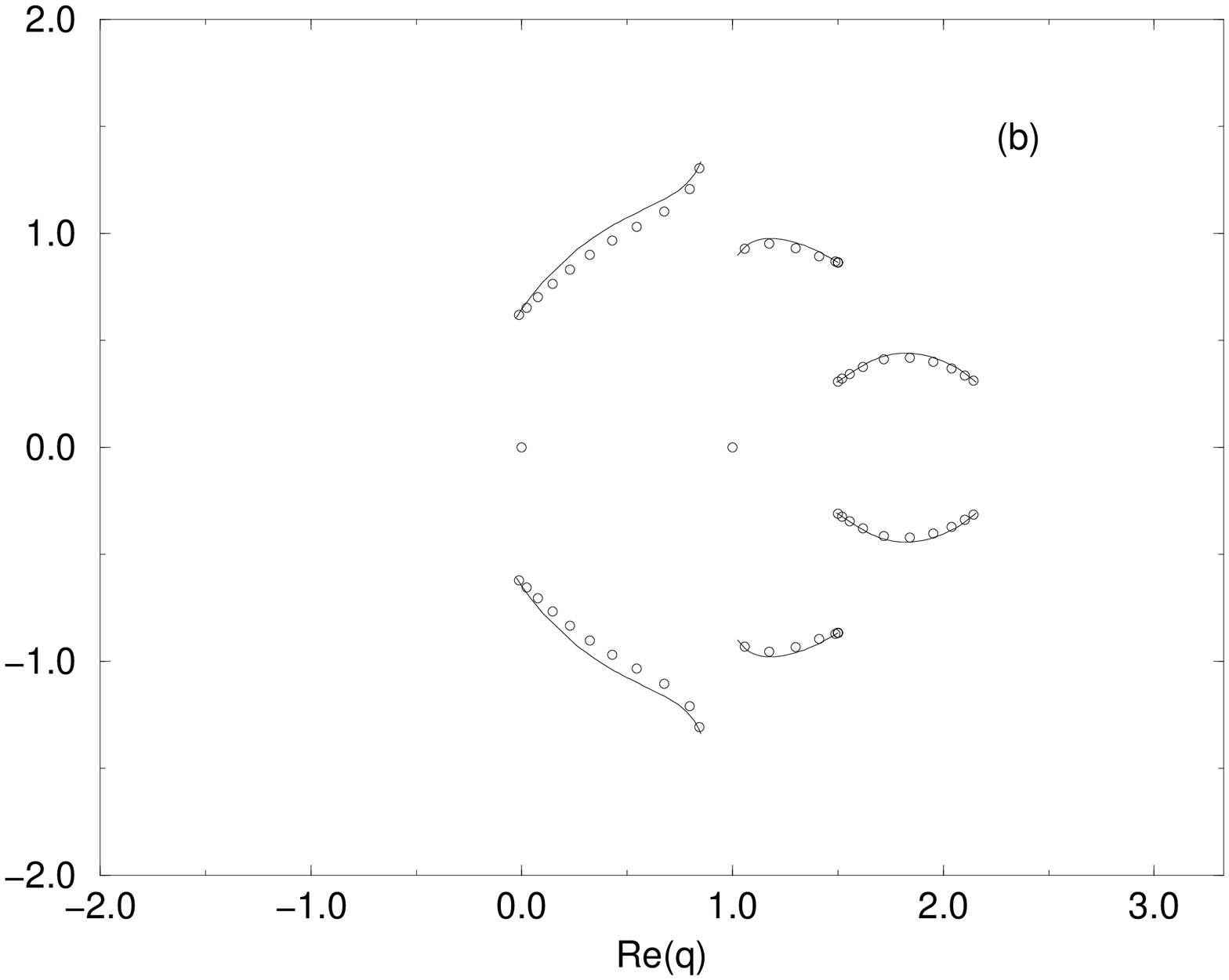}
\end{center}
\vspace{-2cm}
\caption{\footnotesize{As in Fig. \ref{hep3k34} for $p=6$ and $k=$ (a) 3 (b)
4.  For comparison, chromatic zeros are shown for $m=10$.}}
\label{hep6k34}
\end{figure}

\begin{figure}
\vspace{-4cm}
\centering
\leavevmode
\epsfxsize=3.0in
\begin{center}
\leavevmode
\epsffile{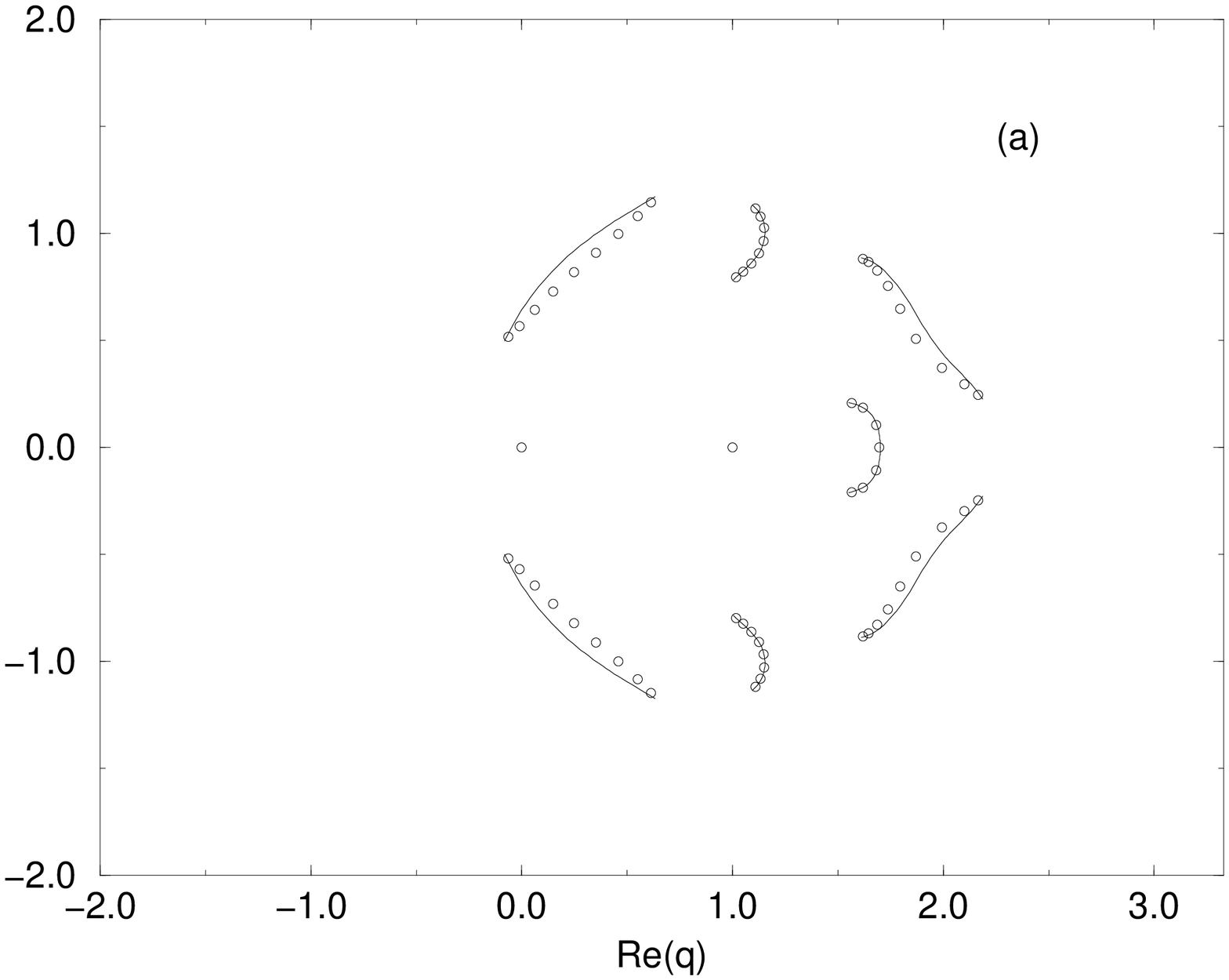}
\end{center}
\vspace{-3cm}
\begin{center}
\leavevmode
\epsfxsize=3.0in
\epsffile{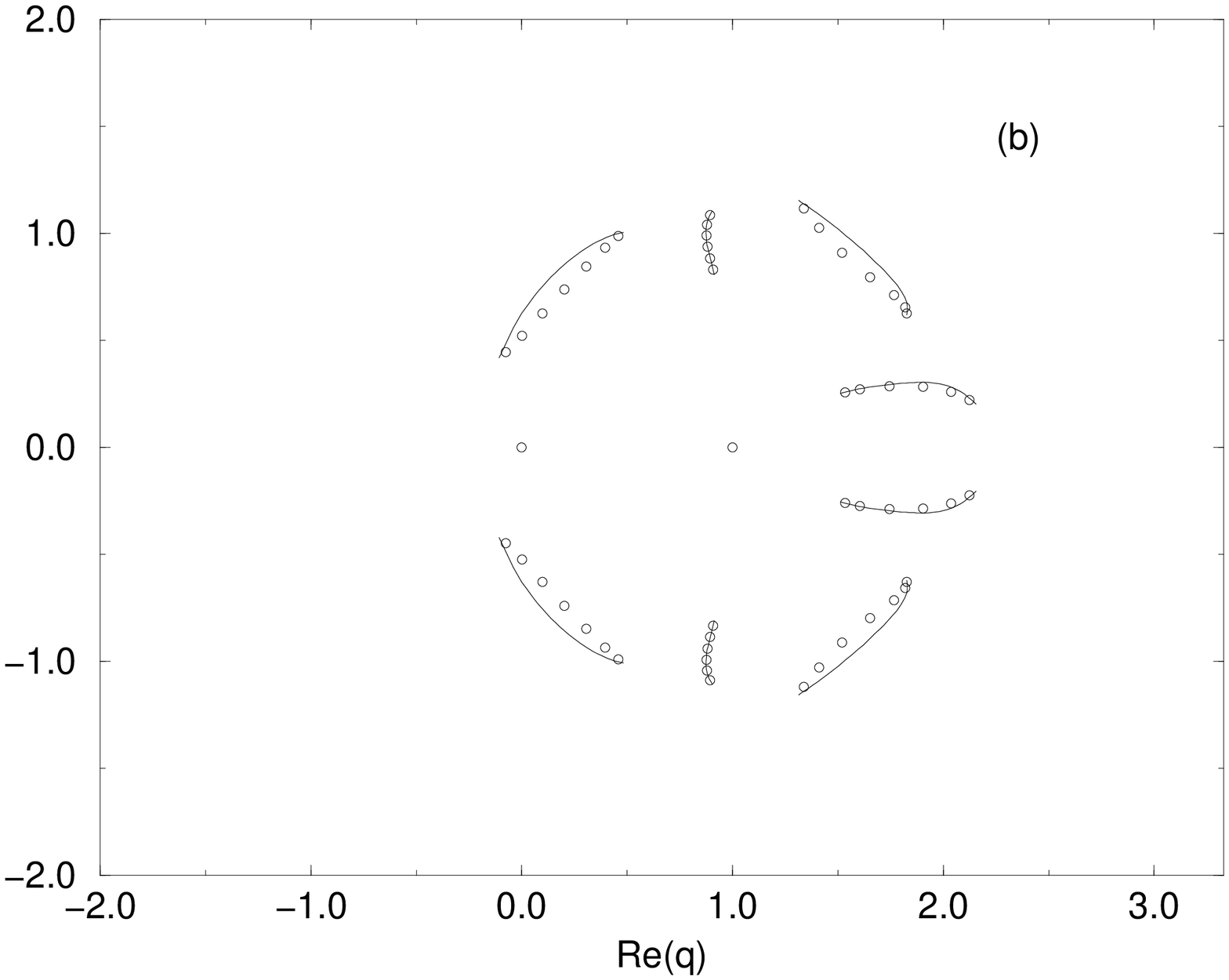}
\end{center}
\vspace{-2cm}
\caption{\footnotesize{As in Fig. \ref{hep3k34} for $p=6$ and $k=$ (a) 5 (b)
6.  For comparison, chromatic zeros are shown for $m=$ (a) 8 (b) 6.}} 
\label{hep6k56}
\end{figure}

As simple examples of the cases of (i) fixed $k, m$ and variable $p$ and 
(ii) fixed $p, m$ and variable $k$, one may take, respectively, 
(i) $k=3$ and $m=1$, and (ii) $p=3$, $m=1$.  In these cases, the general 
results (\ref{chpkm1}) and (\ref{pck}) yield
\beq
P((Ch)_{p,k=3,m=1},q) = P(C_{p+2},q) = a^{p+2} + (-1)^pa
\label{pchpk3m1}
\eeq
\beq
P((Ch)_{p=3,k,m=1},q) = P(C_{2k-1},q) = a^{2k-1} + (-1)^{2k-1}a
\label{pch12km1}
\eeq
where $a=q-1$. One sees the results (\ref{blklp}), (\ref{wr1blklp}),
and (\ref{wr2blklp}) immediately.  For these cases, as we showed
previously \cite{w,wc}, the zeros of the chromatic polynomial lie precisely on
the asymptotic locus (\ref{blklp}) even for finite $p$.  
As a more complicated and generic example of category (ii), fixed $p$, $m$,
and variable $k$, we take $p=3$ and $m=2$. 
In Fig. \ref{hep3k25m2q} we show the zeros of this chromatic polynomial for
$k=25$, in comparison with the asymptotic locus ${\cal B}$ given by eq. 
(\ref{blklp}).  One sees that the chromatic zeros lie close to ${\cal B}$. 

\pagebreak

It is interesting to inquire whether, if we first take $m \to \infty$ with
fixed $p$ and $k$ to obtain ${\cal B}_{L_m}$, and then take, say, $p \to
\infty$, the resulting locus is equal or unequal to that which we obtain by
first taking $p\to \infty$ for fixed $k$ and $m$, thereby getting 
${\cal B}_{L_p}$, and then taking $m \to \infty$.  Indeed, one can formulate
three such questions: 

\vspace{4mm}

\begin{enumerate}

\item

Is $\lim_{k \to \infty}{\cal B}_{G; \ L_m}$ equal or unequal to 
$\lim_{m \to \infty}{\cal B}_{G; \ L_k}$ for fixed $p$ \ ?

\item

Is $\lim_{p \to \infty}{\cal B}_{G; \ L_m}$ equal or unequal to 
$ \lim_{m \to \infty}{\cal B}_{G; \ L_p}$ for fixed $k$ \ ? 

\item

Is $\lim_{p \to \infty}{\cal B}_{G; \ L_k}$ equal or unequal to 
$ \lim_{k \to \infty}{\cal B}_{G; \ L_p}$ for fixed $m$ \ ? 

\end{enumerate}

As discussed above, we shall restrict ourselves here to the nontrivial cases $m
\ge 1$ and $k \ge 3$, and, since we have already discussed the noncompact case
where $p=2$ for the $L_m$ limit, we also restrict to $p \ge 3$ for this $L_m$
limit.  We first observe that from eq. (\ref{blklp}), it follows immediately 
that the two limits in item (3) above do commute. 
As regards the $k \to \infty$ limit of ${\cal B}_{L_m}$, our results in the
previous section show that for odd $p$, ${\cal B}_{L_m}$ includes the line
segment (\ref{linesegmentkinf}) in this limit, and therefore the limits in 
item (1) above are definitely not equal.  Concerning the limits in (2) above, 
although some of the arcs forming ${\cal B}$ in the $L_m$ limit approach the 
unit circle (\ref{blklp}) as $p \to \infty$, others appear to be roughly
orthogonal to this circle, so the limits in (2) could only be equal if these
roughly orthogonally oriented arcs shrink and disappear as $p \to \infty$. 
Thus, the families of graphs studied here exhibit both noncommuting limits 
of nonanalytic loci as certain parameters are taken to infinity in different 
orders, and an example of commuting limits (indeed, in this case, the two 
loci are identical, as noted in eq. (\ref{blklp}).

\section{Conclusions}

In this paper, using a generating function method, we have presented exact 
calculations of the chromatic polynomials for a variety of homeomorphic
families of strip graphs, $(Ch)_{k_1,k_2,\Sigma,k,m}$.  We have studied each of
the ways of sending the total number of vertices $n \to \infty$ and have
determined the respective loci ${\cal B}$ where $W$ is nonanalytic.  In 
particular, we have studied the case of infinitely long strips, i.e., 
$m \to \infty$, with fixed $k_1$, $k_2$, $\Sigma$ (of which only the
combination $p=k_1+k_2$ affects ${\cal B}$), and $k$.  Our results exhibit
considerable richness and complexity in the forms of ${\cal B}$.  In the
future, it would be of interest to continue the calculations of chromatic
polynomials, $W$ functions, and the the loci ${\cal B}$ for other types of
homeomorphic families of graphs.  Our findings yield further insights
into the behavior of the nonzero ground state entropy of $q$-state
Potts antiferromagnets on various graphs. 

\vspace{8mm}

This research was supported in part by the NSF grant PHY-97-22101. 

\pagebreak

\section{Appendix 1}

In this Appendix we gather together some convenient formulas concerning the
function $D_k$, defined in eq. (\ref{dk}) in the text.  From the definition, it
follows that $k \ge 1$.  Some special values are $D_1=0$, \ $D_2=1$, \ and 
$D_3=q-2$.  The degree, as a polynomial in $q$, of $D_k$ is clearly 
\beq
deg_q (D_k) = k-2 \quad {\rm for} \quad k \ge 2 
\label{degreedk}
\eeq
In general, one has 
\beq
D_k(q=0) = (-1)^k (k-1)
\label{dk0}
\eeq
and
\beq
D_k(q=1)=(-1)^k \quad {\rm for} \quad k \ge 2
\label{dk1}
\eeq
Since for $k$ even, the circuit graph $C_k$ is bipartite, which is equivalent
to the fact that the chromatic number $\chi(C_{k \ even})=2$, and thus
$P(C_{k \ even},q=2)=2$, it follows that
\beq
D_{k \ even}(q=2)=1
\label{dkevenq2}
\eeq
Since for $k$ odd, $\chi(C_{k \ odd})=3$ and $P(C_{k \ odd},q=2)=0$, we have
\beq
D_{k \ odd}(q=2)=0
\label{dkoddq2}
\eeq
This zero results from a linear factor, i.e.,
\beq
D_{k \ odd}=(q-2)pol(q)
\label{dkkoddfactor}
\eeq
where $pol(q)$ is a polynomial of degree $k-3$ in $q$ with $pol(q=2) \ne 0$.
Further, for $p \ge 3$ and $k \ge 2$, 
\beq
D_k-aD_{k+2n}=(q-2)pol'(q)
\label{dk2shift}
\eeq
where $pol'(q)$ is another polynomial. 
Some identities that we have derived and used for our calculations
are listed below:
\beq
D_k-aD_{k-1}=(-1)^k
\label{dkdkm1rel}
\eeq
or equivalently, $D_{k+1} - D_k = D_3D_k + (-1)^{k+1}$, and 
\beq
D_{2k-1}=[qD_k-2(-1)^k]D_k
\label{d2km1rel}
\eeq
We note some further important properties.  If $k \ge 4$ is even, 
then the derivative $dD_k/dq$ vanishes at a single real value
of $q$, which we denote $q_m$, and $d^2 D_k/dq^2|_{q=q_m} > 0$, so 
that this is an absolute mininum of $D_{k \ even}$ (the subscript $m$ on 
$q_m$ refers to this \underline minimum).  Note that the restriction on $k$ 
is imposed since $D_2=1$ so $dD_2/dq=0$.  Furthermore, 
\beq
D_{k \ even}(q = q_m) > 0
\label{dkevenminpos}
\eeq
so that 
\beq
{\rm If} \quad k \ge 4 \quad {\rm is \ even, \ then} \quad D_k > 0 
\quad \forall \ q \in {\mathbb R}
\label{dkevenpos}
\eeq
We find that $q_m$ increases monotonically from 
\beq
q_m(k=4)=\frac{3}{2}
\label{qmk4}
\eeq
approaching
\beq
\lim_{k \to \infty} q_m = 2
\label{qmkinf}
\eeq
\beq
{\rm If} \quad k \ge 3 \quad {\rm is \ odd, \ then} \quad 
\frac{dD_k}{dq} > 0 \quad \forall \ q \in {\mathbb R} 
\label{dkevenslope}
\eeq
so that in this case $D_k$ has no critical points (zeros of $dD_k/dq$).
Together with eq. (\ref{dkoddq2}), this implies that if 
$k$ is odd and $k \ge 3$, then $D_k$ has a single real zero, at $q=2$. 
The proofs follow immediately from the definition of $D_k$.

\section{Appendix 2}

In this appendix we show exact calculations of the respective continuous 
nonanalytic loci ${\cal B}$ are shown for two higher values of $p$, viz, 
$p=5$ and $p=6$, and for each $p$, several values of $k \ge 3$.  These results
are discussed in the text. 

\pagebreak

\pagebreak

\pagebreak

\pagebreak

\vfill
\eject

\begin{thebibliography}{99}
\bibitem{graphdef}{We recall the strict mathematical definition of a graph as
a finite collection of vertices and bonds connecting (all or a subset of the)
vertices, such that there are (i) no loops, i.e., bonds connecting a
vertex to itself and (ii) no multiple bonds connecting a given pair of
vertices.}
\bibitem{early}{Some related earlier work is in Refs. 
\cite{bds}-\cite{read91}.}
\bibitem{label}{In Ref. \cite{strip}, the labelling convention was chosen so 
that the total number of graphical subunits was $m+1$; in this paper it is 
chosen so that this number is $m$. Also, in Ref. \cite{strip}, we denoted the
coefficient functions in the numerator ${\cal N}(x)$ of the generating function
as $a_j(q)$; here, in order to avoid confusion with our standard notation
\cite{w} for the terms $a_j(q)$ in eq. (\ref{pgsum}), we use capital $A_j(q)$
for the coefficient functions in ${\cal N}(x)$ in eq. (\ref{n}).}
\bibitem{thm}{We recall the statement of the addition-contraction theorem
\cite{rtrev}-\cite{biggsbook}: let $G$ be a graph, and let $v$ and $v'$ be two
non-adjacent vertices in $G$.  Form (i) the graph $G_{add.}$ in which one
adds a bond connecting $v$ and $v'$, and (ii) the graph $G_{contr.}$ in which
one identifies $v$ and $v'$. Then the chromatic polynomial for coloring $G$
with $q$ colors, $P(G,q)$, satisfies $P(G,q) = P(G_{add.},q) +
P(G_{contr.},q)$. The reverse process of bond deletion, leading to the same 
equation, is known as the deletion-contraction theorem.}
\bibitem{complete}{The complete graph $K_n$ with $n$ vertices is defined as the
graph where each of these vertices is connected to all of the other $n-1$
vertices by bonds.  The complement $\overline K_n$ of $K_n$ is the graph with 
$n$ vertices and no bonds. A tree graph $T_n$ is an $n$--vertex graph
containing no circuits. The ``join'' of two graphs $G$ and $H$ is obtained
by adding bonds connecting each of the vertices of $G$ to those of $H$.  This
was denoted $G \times H$ in Ref. \cite{wa}; in Ref \cite{wa23} and here we use
the more common notation for this object in the mathematical literature, viz.,
$G+H$.}
\bibitem{chromatic}{The chromatic number $\chi(G)$ of a graph $G$ is defined as
the minimum number of colors, $q \in {\mathbb Z}_+$ which are necessary to 
color the graph such that no two adjacent vertices have the same color, i.e.,
the minimum $q \in {\mathbb Z}_+$ such that $P(G,q) > 0$. }
\bibitem{global}{Our definition of ``global circuit'' in Ref. \cite{strip} 
was the following: a regular lattice graph has a global circuit if and only 
if by travelling along a lattice vectors one can arrive back at one's
starting point.  Here the definition of a lattice vector is clear for a
Cartesian $d$-dimensional lattice; for the triangular lattices, the
lattice vectors can be taken to lie along the $0^\circ$, $60^\circ$, and
$120^\circ$ directions.  For a honeycomb lattice, drawn as a brick lattice, the
paths can be taken to lie along the brick (horizontal) direction, and the 
square-sawtooth paths in the vertical direction.  
(A generalization of the definition is necessary for
heteropolygonal lattices \cite{ww}, but this is not relevant here.)
For a $d$-dimensional
Cartesian lattice, if one imposes free boundary conditions, there are no global
circuits, while if one imposes periodic boundary conditions in at least one
direction, then there are global circuits. In the thermodynamic limit, the 
lengths of these global circuits go to infinity.  If one considers an 
$L_x \times L_y$ strip and only takes $L_x \to \infty$, then one must 
distinguish between two kinds of possible circuits: (i) if one imposes 
periodic boundary conditions in the longitudinal ($x$) direction, then there 
are global circuits in the sense intended in Ref. \cite{strip}, whose length 
$L_x \to \infty$; (ii) if one 
imposes periodic boundary conditions in the transverse ($y$) direction, then 
there are what could be called local circuits along the $y$ direction, with 
$L_y$ remaining finite.}
\bibitem{pol}{For the case $p=2$, $R$ is a perfect square: 
$R=(D_k-(q-1)D_{k-1})^2$, so that $\lambda_{1,2}$ are polynomial rather than 
algebraic and hence have no branch point singularities. 
For the case $k=2$, i.e., no tranverse homeomorphic expansion, 
$b_2=0$ (and $A_1=0$), and there is only one $\lambda=D_p$, which again is
polynomial; ${\cal D}=(1-D_px)$.}
\end{thebibliography}
\end{document}